\documentclass[11pt,a4paper]{article}
\usepackage{geometry}  
\usepackage[british]{babel} 
\usepackage{a4wide}
\usepackage[T1]{fontenc}
\usepackage[utf8]{inputenc}
\usepackage[pdftex]{graphicx}  
\usepackage{charter} 
\usepackage{varioref}\labelformat{equation}{(#1)}
\usepackage[round]{natbib}
\usepackage{enumerate,mathrsfs,url}
\usepackage{dcolumn} 
\newcolumntype{d}{D{.}{.}{2.5}}
\newcolumntype{s}{D{.}{.}{1.2}}
\usepackage{amsfonts}
\newcommand{\anymode}[1]{\ifmmode{#1}\else\mbox{$#1$}\fi}

\newcommand{\NB}[1]{\texttt{/\textsuperscript{NB}}%
          \marginpar{\raggedright\sl\small#1\hfill}}	  

\long\def\Nota#1{\footnote{#1}\kern-0.2em\NB{cfr Nota$^{(\thefootnote)}$}}
\def\linea#1{\ifhmode\hfill\break\fi\hbox to \hsize{#1}}

\newcommand{\inv}{^{-1}}

\def\vc{v\kern -0.1em .c.\relax}

\newcommand{\dfrac}[2]{\displaystyle{\frac{#1}{#2}}}

\newcommand{\E}[2][]{
   \ensuremath{\mathbb{E}_{#1}\!\left\{\displaystyle{#2}\right\}}}

\newcommand{\half}{\mbox{$\textstyle \frac{1}{2}$}}   % p.172
\long\def\ignore#1{}

\newcommand{\indep}{\perp\kern-0.5em\perp}

\newcommand{\pr}[2][]{
   \ensuremath{\mathbb{P}_{#1}\!\left\{\displaystyle{#2}\right\}}}

\newcommand{\Real}{\mathbb{R}}

\newcommand{\tended}{\hbox{$~\stackrel{d}{\longrightarrow}~$}}

\newcommand{\T}{^{\top}}
\newcommand{\var}[2][]{
   \ensuremath{\textrm{var}_{#1}\!\left\{\displaystyle{#2}\right\}}}

\newcommand{\vech}{\mathop{\mathrm{vech}}\nolimits}

\newcommand{\SN}{\mathrm{SN}{}}
\newcommand{\ST}{\mathrm{ST}{}}
\newcommand{\N}{\mathrm{N}{}}

\title{\bf Maximum penalized likelihood estimation \\
    for skew-normal and skew-$t$ distributions }
\author{
  {\Large Adelchi Azzalini} \\  \large 
  Dipartimento di Scienze Statistiche \\
  Università di Padova\\
  Italia 
  \and
  {\Large Reinaldo B.\ Arellano-Valle} \\
  Departamento de Estadística \\
  Pontificia Universidad Católica de Chile  \\
  Santiago,  Chile
 }
\date{\footnotesize\today}
\begin{document}
\maketitle
\begin{abstract}
  The skew-normal and the skew-$t$ distributions are parametric families which
  are currently under intense investigation since they provide a more
  flexible formulation compared to the classical normal and $t$ distributions
  by introducing a parameter which regulates their skewness.  While these
  families enjoy attractive formal properties from the probability viewpoint,
  a practical problem with their usage in applications is the possibility that
  the maximum likelihood estimate of the parameter which regulates skewness
  diverges.  This situation has vanishing probability for increasing sample
  size, but for finite samples it occurs with non-negligible probability, and
  its occurrence has  unpleasant effects on the inferential process.  
  Methods for overcoming this
  problem have been put forward both in the classical and in the Bayesian
  formulation, but their applicability is restricted to simple situations.
  We formulate a proposal based on the idea of penalized
  likelihood, which has connections with some of the existing methods, but it
  applies more generally, including in the multivariate case.
\end{abstract}  

\vspace{2ex}

\noindent\emph{Some key-words:} anomalies of maximum likelihood estimation, 
boundary estimates, penalized likelihood, skew-elliptical distributions.
% ======================================================
\clearpage
\section{Skew-normal distribution: inferential issues}

\subsection{Background}
A currently active stream of literature deals with a set of probability
distributions whose most prominent representative is the skew-normal
distribution, whose density function in the scalar case is
\begin{equation} \label{e:sn-pdf}
  \frac{2}{\omega}\:\phi\left(\frac{x-\xi}{\omega}\right)\:
     \Phi\left(\alpha\:\frac{x-\xi}{\omega}\right), \qquad x\in\Real,
\end{equation} 
where $\phi$ and $\Phi$ denote the $\N(0,1)$ density and distribution
function, respectively. The skew-normal density depends on parameters $\xi$,
$\omega$ (with $\omega>0$) and $\alpha$, which regulate location, scale and
shape, respectively.  If $Y$ is a random variable with density function
\ref{e:sn-pdf}, we shall write $Y\sim\SN(\xi,\omega^2,\alpha)$.  When
$\alpha=0$, we return to the regular normal distribution $\N(\xi,\omega^2)$;
otherwise the distribution is positively or negatively asymmetric, in agreement
with the sign of $\alpha$.
 
The basic construction \ref{e:sn-pdf} can be extended in several directions,
to various levels of generality, leading to a much broader set of
distributions; the terms skew-elliptical and skew-symmetric distributions are
usually adopted in this context.  We shall introduce some of these other
constructions in the course of the paper where appropriate, specifically its
multivariate version and the closely-related skew-$t$ distribution. For a
general overview of the subject, we refer the readers to the book edited by
\citet{genton:2004-SE} and the review paper of \citet{azzalini:2005}; a
concise account on the skew-normal distribution, including its multivariate
version, is given by \citet{azzalini:2011-iess-sn}.

Much of the appeal of distribution \ref{e:sn-pdf} comes from its mathematical
tractability and from a number of  formal properties which either replicate
or at least resemble those of the normal distribution, so that they support
the adoption of the name `skew-normal'.  These properties are discussed at
length in the above-quoted references and we do not dwell into this aspect
which is outside the scope of the present paper.

The statistical side of the treatment of \ref{e:sn-pdf} shows instead two
peculiar features which call for special treatment if one wants to use this
distribution in data analysis.  Given a random sample with components
independently drawn from \ref{e:sn-pdf}, the first of these problematic
aspects refers to the specific value $\alpha=0$, and it shows up in a few
intimately related manifestations, all originated by the proportionality of
the score functions for $\xi$ and $\alpha$ to each other.  The main
implications of this fact are that, at $\alpha=0$, (i)~for any sample, the
profile log-likelihood function for $\alpha$ has an inflection point, (ii)~the
expected information matrix is singular, even if the distribution is
identifiable.

This singularity issue has given rise to much concern, being often perceived
as a major structural problem of the skew-normal family of distributions,
while it is only a problem of the adopted parameterization.  Moving from
$(\xi, \omega, \alpha)$ to the `centred parameterization' proposed by
\citet{azzalini:1985}, essentially the cumulants up to the third order with
the third one in standardized form, removes all these issues.  For a more
extended discussion of this point and for other relevant references, see
\S\,2.4 of \citet{azzalini:2005}. For a multivariate version of the centred
parameterization, see \citet{arel:azza:2008}.

%------------
\subsection{MLE boundary values}

The present paper deals instead with the second one of the two peculiar
aspects mentioned above, represented by the fact that, with non-zero
probability, the maximum likelihood estimate (MLE) of $\alpha$ diverges.
The problem is easily examined in the one-parameter case $\SN(0,1,\alpha)$
where the log-likelihood based on a random sample $z=(z_1,\ldots, z_n)$ is
\begin{equation}  \label{e:logL-sn1}
  \ell(\alpha)= \mathrm{constant} + \sum_{i=1}^n \zeta_0(\alpha\, z_i)
\end{equation}
where $\zeta_0(x)=\log\{2\:\Phi(x)\}$. Since $\zeta_0$ is a monotonically
increasing function, it is then immediate that the maximum of $\ell$ is at
$\alpha=\infty$ when all $z_i>0$, and it is at $\alpha=-\infty$ if all
$z_i<0$, as noted by \citet{liseo:1990}. Therefore, if the data have all equal
sign, their actual location is irrelevant. The value $\alpha=\infty$
corresponds to the half-normal or $\chi$ distribution; if $\alpha=-\infty$ the
$\chi$ distribution is mirrored on the negative axis.

Further, it is only when all sample values have the same sign that we get a
divergent MLE, since it can be shown that, 
when there are observations with opposite sign, the MLE is finite 
\citep{mart:vare:etal:2008}. 

Taking into account the known fact $\pr{Z<0}=\half+\pi\inv\arctan\alpha$, 
when $Z\sim \SN(0,1,\alpha)$,  the probability of a divergent MLE is 
immediately written as
\[
    p_{n,\alpha} = 
        \left(\frac{1}{2}-\frac{\arctan\alpha}{\pi}\right)^n+
        \left(\frac{1}{2}+\frac{\arctan\alpha}{\pi}\right)^n \,.
\] 
This probability goes rapidly to $0$ as $n\to\infty$, provided
$|\alpha|<\infty$, but for small or moderate sample size it can be
non-negligible, especially if $\alpha$ is far from 0. 
To get an idea, consider that $p_{25,5}\approx 0.197$ and 
$p_{50,5}\approx 0.039$.

In the three-parameter case $\SN(\xi, \omega^2, \alpha)$, infinite values of
the MLE can occur as well, but a characterization of the samples leading to
such estimates has not been obtained, as far as we know.  It is convenient to
illustrate this case with the aid of a numerical example, and we make use the
so-called `frontier data', presented by \citet{azza:capi:1999}, which is a set
of $n=50$ values sampled from $\SN(0,1,5)$. For these data, the MLE 
$\hat\alpha$ of $\alpha$  diverges when one assumes a three-parameter 
$\SN(\xi, \omega^2, \alpha)$ family of distributions.

\begin{figure}
\centerline{
  \includegraphics[width=0.49\hsize]{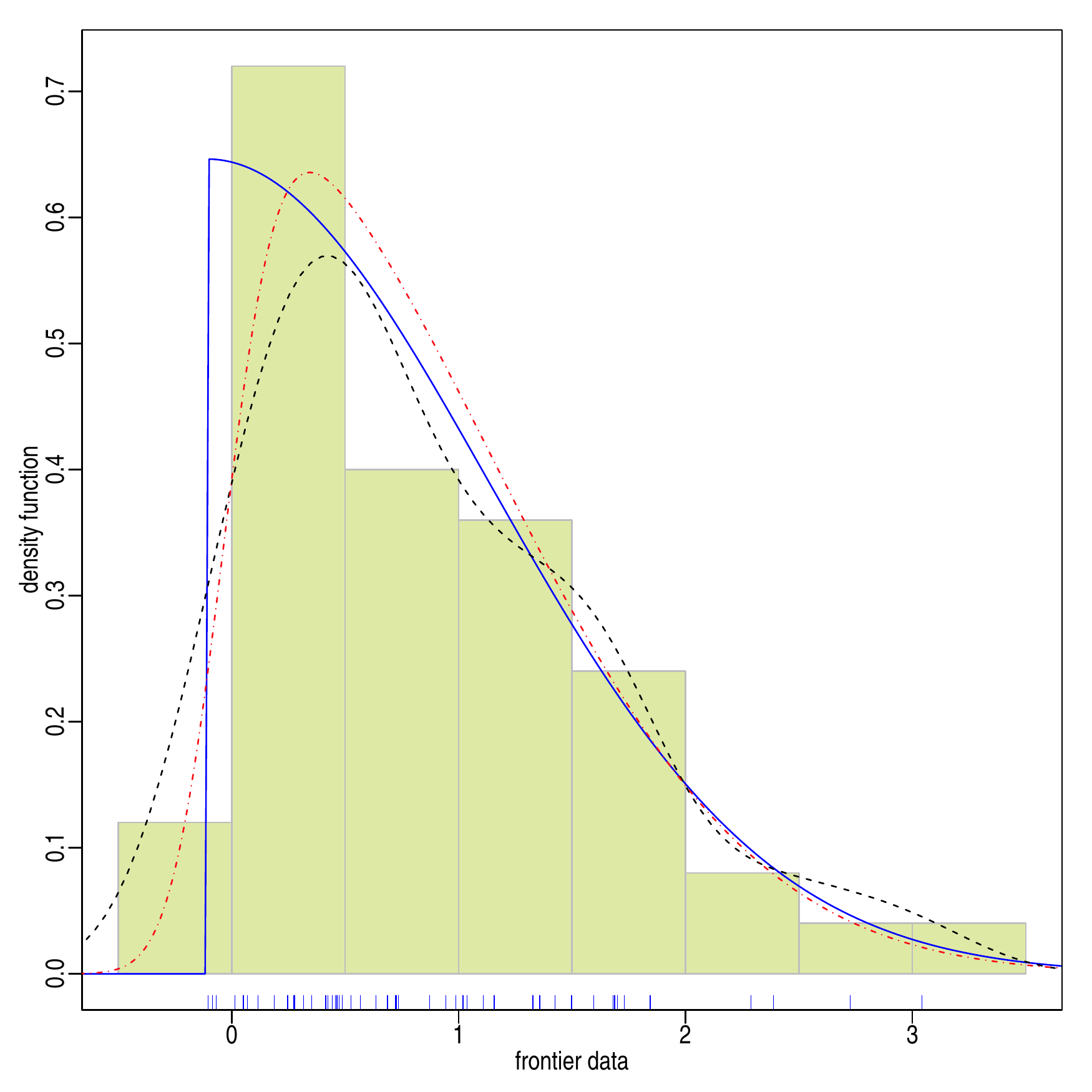}\hfill
  \includegraphics[width=0.49\hsize]{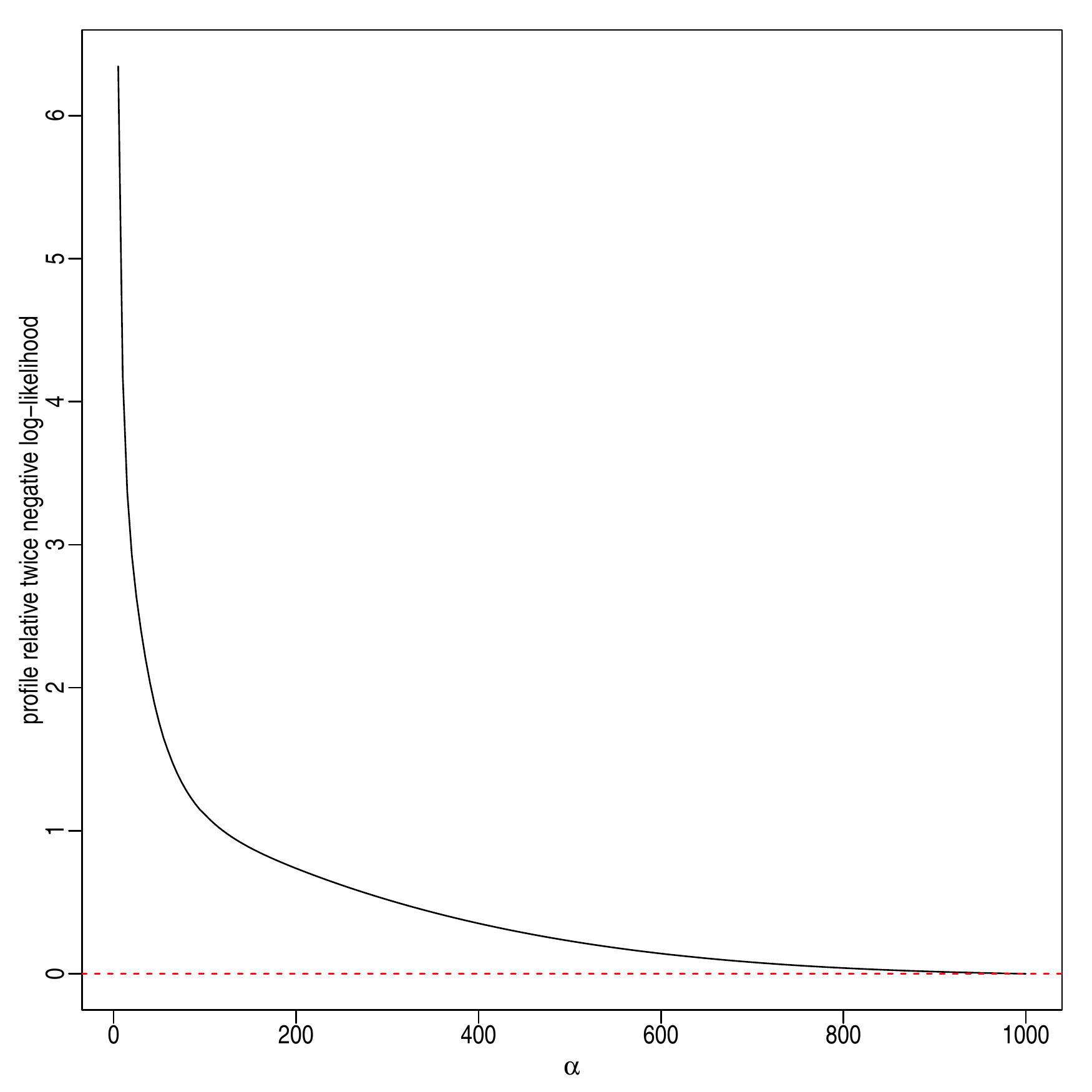}}
\caption{\textsl{ Frontier data. Left panel: rug-plot and histogram with
    superimposed MLE fit (continuous line), non-parametric fit (dashed line)
    and MPLE fit (dot-dashed line). Right panel: deviance function of
    $\alpha$. }}
\label{f:frontier-data}
\end{figure}

The left panel of Figure~\ref{f:frontier-data} displays these data together
with their histogram and two fitted curves: one corresponds to the MLE,
another one is a non-parametric kernel-type estimate, using a Gaussian kernel
with bandwidth chosen by cross-validation, and the third curve will be
described later on.  Since  $\hat\alpha=\infty$, the
latter curve is a shifted and scaled $\chi$ distribution, with origin just
below the smallest sample value.  The right-side panel of this figure displays
the deviance function
\[
  D(\alpha) = 2\:\{\log L^*(\hat\alpha)- \log L^*(\alpha)\}
\]  
where $L^*(\alpha)$ denotes the profile likelihood for $\alpha$.  The curve,
which appears to be monotonically decreasing, becomes flat for large
$\alpha$.

As mentioned earlier, the inclusion of the limiting points $\alpha=\pm\infty$
in the parameter space is admissible for distribution \ref{e:sn-pdf}.
However, $\alpha=\pm\infty$ represent a peculiar situation, not only because
we are at the boundary of the parameter space, but in addition the support of
the distribution collapses to the half-line, instead of the complete real line
as for any finite $\alpha$.
  
When an unbounded estimate, $\hat\alpha=\pm\infty$,  occurs 
there are two alternative aptitudes of a statistician. One is to say: 
if the MLE is $\hat\alpha=\infty$, we
still take it; after all, this is an admissible value of the parameter.
Notice however that, on the boundary on the parameter space, standard
asymptotic distribution theory of MLE does not hold, and a special theory must
be developed to obtain standard errors of the estimates.  The other aptitude
is to disregard $\hat\alpha=\infty$ as an anomaly of MLE. Not only this
parameter point is peculiar for the general reasons indicated earlier, but in
addition it often does not appear to actually describe the data in
a satisfactory way.
For instance, in the  case of Figure~\ref{f:frontier-data}, neither 
the histogram nor the non-parametric density estimate
exhibit the extreme pattern in the data which are implied by the MLE value.
Furthermore,  as remarked by \citet{azza:capi:1999}, the sample index of
skewness of the data, $0.902$, is well inside the admissible range of
$\gamma_1$, about $\pm0.99527$, whose extremal values correspond to
$\alpha=\pm\infty$.
 
\begin{figure}
\centerline{
  \includegraphics[width=0.48\hsize]{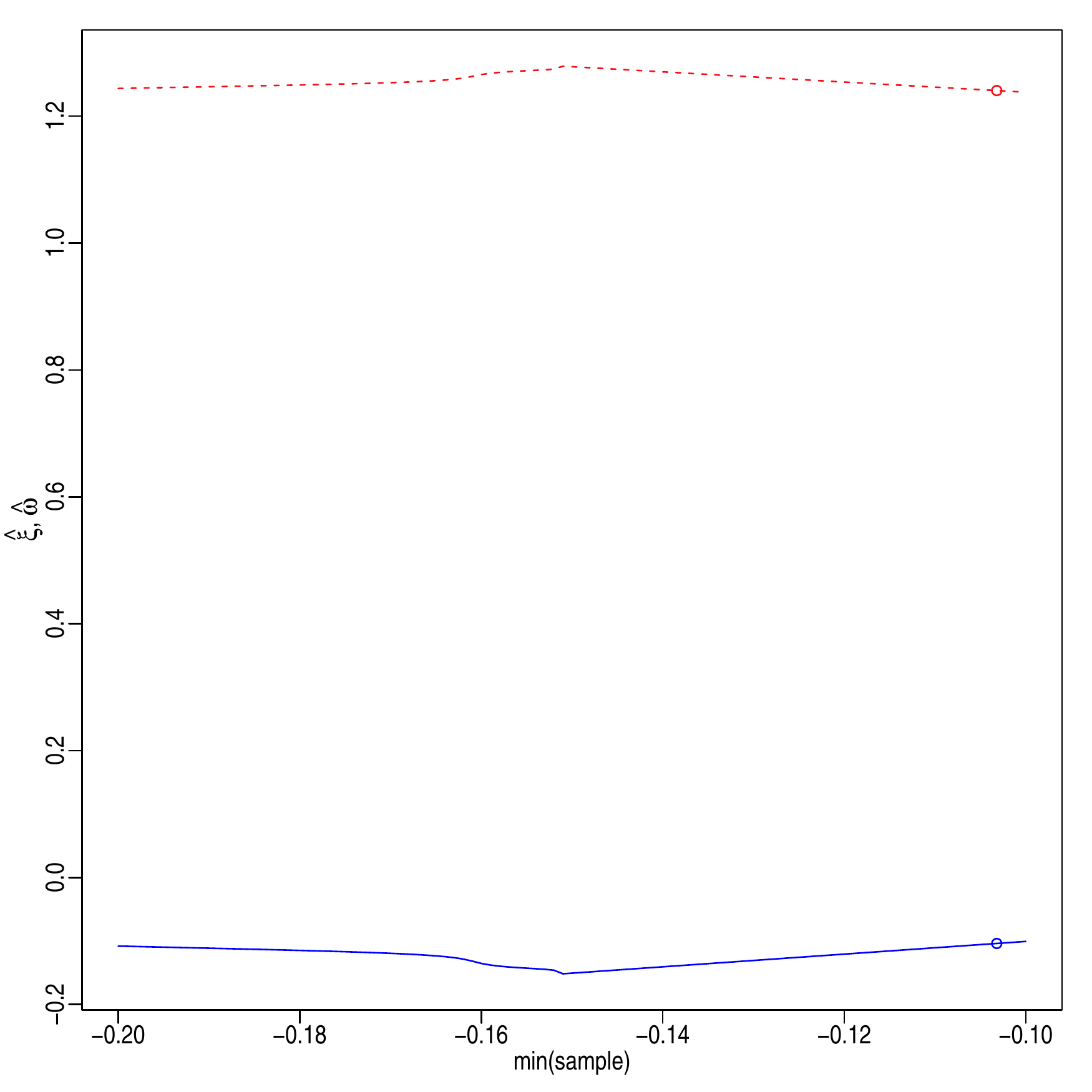}\hfill
  \includegraphics[width=0.48\hsize]{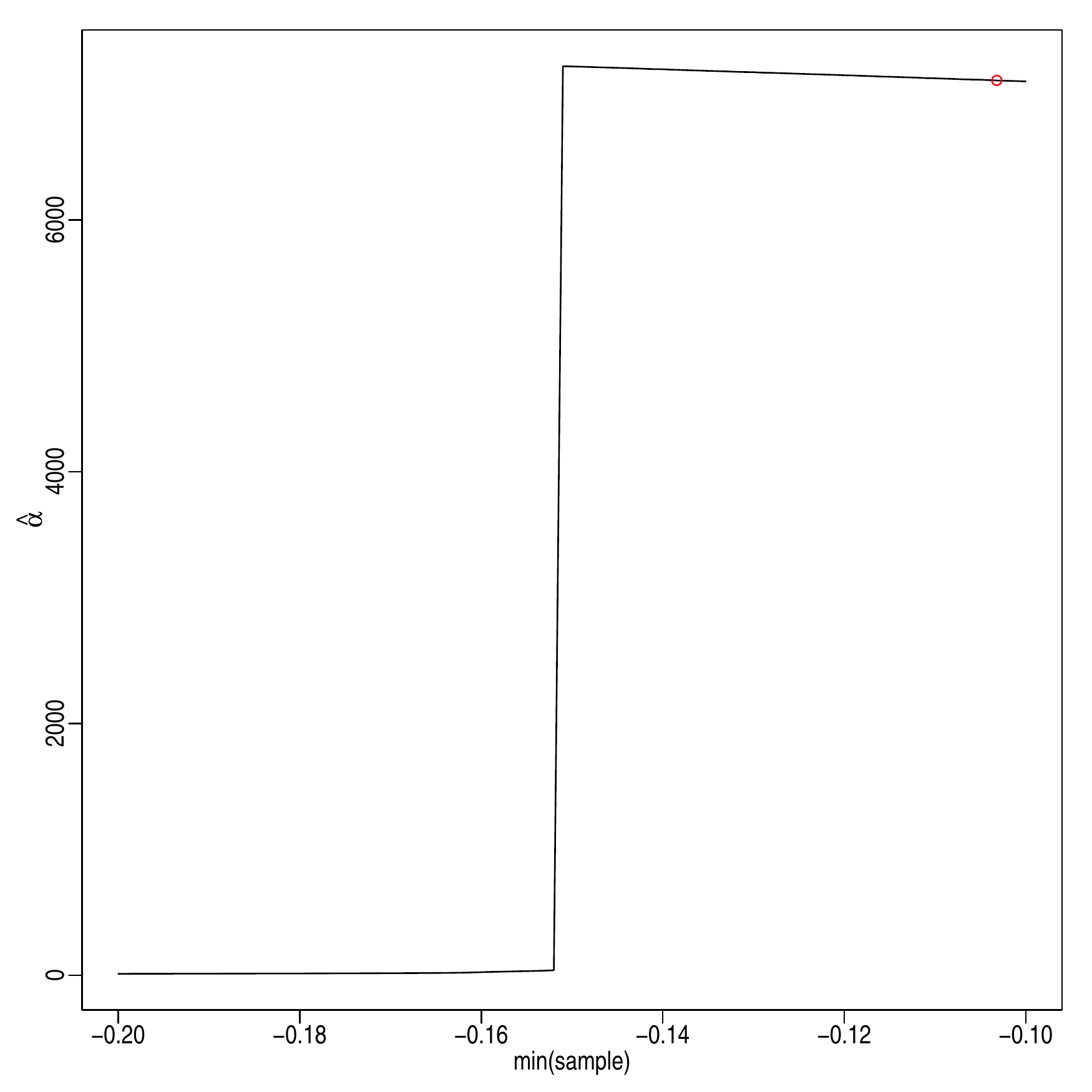}}
\caption{\textsl{
  Frontier data: evolution of the MLE components when the minimum 
  sample value ranges $-0.20$ to $-0.10$. The left panel refers to 
  $\hat\xi$ (bottom curve) and $\hat\omega$ (top curve); the right 
  panel refers to $\hat\alpha$. The bullets denote the estimates
  using the original sample minimum.}}
\label{f:frontier-mle}
\end{figure}

Yet another argument against the MLE choice is provided by the plots in
Figure~\ref{f:frontier-mle}, which displays the behaviour of the three MLE
components when the minimum sample value, $-0.1032$, is replaced by another
value, $m$ say, which ranges from $-0.20$ to $-0.10$. While in the left panel
$\hat\xi$ and $\hat\omega$ are very stable as $m$ moves along the range, the
evolution of $\hat\alpha$ in the right panel has a dramatic discontinuity.  
When $m$ varies from $-0.2$ to $-0.152$, $\hat\alpha$ increases gradually from
about $12$ to about $40$, but at $m=-0.151$ it jumps above $7200$. This value
is however only where the numerical optimization procedure was stopped
searching, but the search would lead to increasingly large values if it was
left running, although the divergence of $\hat\alpha$ corresponds to a 
negligible increase of the log-likelihood function, as indicated by the right
panel of Figure~\ref{f:frontier-data}.  Such a severe instability of an
estimator in reaction to this minute variation of a single sample value is
unacceptable on general grounds.

%------------
\subsection{Alternative options}

There appear to exist both general arguments and numerical evidence against
the MLE solution, at least when it leads to boundary values of $\hat\alpha$.
Notice that the problem cannot be cured by reparameterization, like for the
singular information matrix, since any regular transformation maps 
boundary points of the parameter space $(\xi,\omega,\alpha)$ into boundary
points of the new parameter space. In addition, if $\hat\alpha=\pm\infty$, the
equivariance property of MLE would lead to take the transformed boundary point
as the new MLE. Therefore a different estimation method need to be considered.

A number of alternative proposals have been put forward, adopting a range of
different approaches. A preliminary solution to the problem has been put
forward by \cite{azza:capi:1999} in the discussion following the presentation
of the frontier data. This is based on the consideration that the
log-likelihood function varies little over a large span of the $\alpha$ axis;
see the right panel of Figure~\ref{f:frontier-data} for a visual perception at
leas of the profile version of the log-likelihood. It is then reasonable to
take a value whose log-likelihood is below the maximum by a non-significant
amount. While this technique works well in practice and it can applied also to
a variety of similar problems, it leaves some arbitrary margin on the choice of
the acceptable amount of drop from the maximum.

\citet{sartori:2006} has specialized the general bias-reduction method of 
\citet{firth:1993} to the present context.   This technique replaces 
the usual  likelihood equation $\ell'(\alpha)=0$ by the modified form
\begin{equation} \label{e:SF-eqn}
                 \ell'(\alpha) + M(\alpha)= 0
\end{equation}
and the correction term $M(\alpha)$ in the case $Z\sim\SN(0,1,\alpha)$  
takes the form
\begin{equation} \label{e:M,a.p}
    M(\alpha) = -\frac{\alpha}{2}\:\frac{a_4(\alpha)}{a_2(\alpha)},
    \qquad
       a_{p}(\alpha) = \E{Z^p\: \zeta_1(\alpha\,Z)^2}
\end{equation}
where $\zeta_1(x)=\zeta_0'(x)$ is the inverse Mills ratio.  Sartori shows
that, for any sample, the modified likelihood equation has at least one finite
solution. An interesting feature is the close similarity of the shape of
$M(\alpha)$ with the derivative of the logarithm of Jeffreys' uninformative
prior.
	Since the three-parameter case $\SN(\xi,\omega^2,\alpha)$ is hard to
	tackle via the general Firth's method, Sartori introduces a specifically
	constructed two-step scheme.
%where the location and scale parameters are first 
%estimated by MLE,  reducing the problem to the one-parameter case.
%Next, the score function $\ell'(\alpha)$ in \ref{e:SF-eqn} is replaced
%by the profile score function for $\alpha$; the existence of a finite root
%of the estimating equation is still gauranteed.
%Here $a_2$ represents the expected Fisher information for $\alpha$ 
%provided by a single observation. 

In a Bayesian framework, \citet{lise:lope:2006} adopt the Jeffreys prior
for $\alpha$, which they prove to be a proper distribution over the real line.
For the thee-parameter case, an expression of the reference-integrated 
likelihood is obtained, although this is difficult to use for $n$ not small.
Follow-up work has been done by \citet{baye:bran:2007}  whose development 
includes a closed-form approximation
\begin{equation} \label{e:M-BB}
  M(\alpha) \approx 
     -\frac{3\,\alpha}{2}\left(1 +  \frac{8\,\alpha^2}{\pi^2}\right)\inv \,.
\end{equation}
which is based on replacing the normal distribution function entering in the
expression of $a_p(\alpha)$ by a rescaled logistic distribution. The
subsequent simulation study confirms the closeness of the Sartori-Firth
estimate to the Jeffreys' posterior mode.

An alternative route has been taken by \cite{greco:2011} using a minimum
Hellinger distance criterion, which also leads to finite estimates of $\alpha$
for the case $\SN(0,1,\alpha)$. This approach works for the three-parameter
case as well, without introducing special adaptation. However it involves 
the choice of a specific density estimate and of the connected smoothing 
parameter, which influences the final outcome.

The above-recalled constructions lead to elegant results for the basic case
$\SN(0,1,\alpha)$, but the three-parameter case $\SN(\xi,\omega^2,\alpha)$
already poses non-trivial additional difficulties for the first two of them.
The multivariate case has not been tackled at all, as far as we know, except
for a very brief mention of \cite{greco:2011}. The analogous problem with the
skew-normal distribution replaced by the skew-$t$ distribution is inevitably
more complex, because of one additional parameter involved and the diminished
mathematical tractability; we shall review the existing results for the
univariate case in Section~\ref{s:ST}.

The aim of the rest of the paper is to develop a procedure which can be
applied to a range of situations, including the multivariate case, with the
requirement that its behaviour is largely the same of the MLE, with only a
minor modification to prevent boundary estimates.  We first develop our
proposal for the skew-normal distribution, and later extend it to the skew-$t$
distribution.

% ======================================================
\section{Penalization of the log-likelihood function}  \label{s:penalize-logL}

\subsection{General remarks} \label{s:logL.p}

Penalization of the log-likelihood function is a device which has been adopted
in a number of problems to correct some undesirable behaviour of the regular
MLE. \citet[p.\,4262]{sartori:2006} has remarked that \ref{e:SF-eqn} can 
be viewed in this light.

Our aim is to avoid divergent estimates of $\alpha$, in a formulation 
applicable to a wide range of situations of the context described earlier. 
To this end, consider a function of the form
\begin{equation} \label{e:logL.p}
  \ell_p(\theta) = \ell(\theta) - Q 
\end{equation}
where $\ell(\theta)$ denotes the log-likelihood function for 
$\theta$ which denotes the whole set of parameters associated to the
chosen parametric family, and $Q$ represents a non-negative quantity which
penalizes the divergence of $\alpha$ and it remains $O_p(1)$ as $n$ increases.
A value $\tilde\theta$ which maximizes $\ell_p(\theta)$ will be called a
Maximum Penalized Likelihood Estimate (MPLE).

The log-likelihood functions which we have in mind are primarily of
skew-normal type, and related ones discussed in Section~\ref{s:ST}, but part
of the development can potentially be of interest also in other settings.  It
is assumed that $\ell(\theta)$ satisfies the standard conditions for
consistency and asymptotic normality of the regular MLE, $\hat\theta$, as set
for instance in Theorem 5.2.2 of \citet{senPK:sing:1993}. 

Besides the univariate distributions $\SN(0,1,\alpha)$ and
$\SN(\xi,\omega^2,\alpha)$, we consider also the multivariate skew-normal
distribution $\SN_d(\xi,\Omega,\alpha)$ whose density function is
\begin{equation} \label{e:msn-pdf}
  2\:\phi_d(x-\xi; \Omega)\: \Phi\left(\alpha\T\omega\inv(x-\xi)\right),
  \qquad x\in\Real^d\,,
\end{equation}
where $\phi_d(x;\Omega)$ denotes the $\N_d(0,\Omega)$ density function and
$\omega$ is a diagonal matrix formed by the standard deviations of $\Omega$;
in this case $\alpha$ and $\xi$ are $d$-dimensional parameters. For these
three parametric families, the parameter $\theta$ in \ref{e:logL.p} has $1$ or
$3$ or $d(d+5)/2$ components, respectively.

The translation into mathematical notation of the above-indicated requirements
for $Q$ is that
\begin{equation} \label{e:Q-assume}
     Q\ge0 \,, \qquad 
     Q\big|_{\alpha=0}=0 \,, \qquad 
     \lim_{\alpha_j\to\pm\infty} Q =\infty 
\end{equation}
where $\alpha_j$ is the $j$-th component of $\alpha$, for $j=1,\dots,d$. 
For the SN distribution, and the ST distribution to be discussed later,
$\log\,L$ does not diverge to $+\infty$ even when the MLE of $\alpha$
diverges; combining this fact with the third requirement in \ref{e:Q-assume}
we are ensured that \ref{e:logL.p} has a finite maximum in the interior
of the parameter space.
In a different context where some components of the MLE can diverge
but the log-likelihood itself is bounded from above, the same 
argument applies provided the third condition  in \ref{e:Q-assume}
is suitably adapted to the different parameter set.

In the next sections $Q$ will be a function of the parameters only, not
depending on the data. This condition could be removed as long as $Q$ is
$O_p(1)$; however, the mathematical treatment would be more elaborate and for
simplicity we do not consider this case in detail.  For the subsequent
development we also require that $Q$ is twice differentiable with respect to
$\theta$, and that $Q''(\theta)$ is a uniformly continuous mapping in a
neighbourhood of the true parameter point.  An additional sensible
requirement, although not necessary for our construction, is that $Q$
increases monotonically with each $|\alpha_j|$.

Besides existence of $\tilde\theta$, another implication of this formulation
concerns the first-order asymptotic distribution of $\tilde\theta$ when a
random sample of size $n$ is available: as $n\to\infty$, this asymptotic
distribution coincides with the one $\hat\theta$. This fact is intuitive on
noticing that both $\ell(\theta)$ and $\ell_p(\theta)$ are $O_p(n)$ and they
differ by $Q$, which is $O_p(1)$, but it can also easily be proved
formally, under the above regularity conditions, following essentially an
argument similar to Theorem~5.2.2 of \citet{senPK:sing:1993}.  We then
conclude that $\tilde\theta$ is consistent with asymptotic distribution
\begin{equation} \label{e:asympt-mple}
  \sqrt{n}(\tilde\theta -  \theta)\tended \N\left(0, i_E(\theta)\inv\right), 
   \quad\mathrm{when~} n\to\infty\,,
\end{equation}
where $i_E(\theta)$ denotes the expected Fisher information for a single 
observation.  
A more informative expression can be obtained by expanding  
$\ell'_p(\tilde\theta)$ from the point $\hat\theta$ as follows: 
\begin{eqnarray*}
  0 &=& \ell'_p(\tilde\theta)\\
    &=& \ell_p'(\hat\theta) + \ell_p''(\hat\theta) \,(\tilde\theta-\hat\theta)
        + o(\|\tilde\theta-\hat\theta\|)\\
    &=& -Q'(\hat\theta) + \ell_p''(\hat\theta) \,(\tilde\theta-\hat\theta)
        + o(\|\tilde\theta-\hat\theta\|)
\end{eqnarray*}
where  $\ell''_p$ denotes the matrix of second
order derivatives of $\ell_p$.  We can then write
\begin{equation} \label{e:tilde-hat}
   \tilde\theta-\hat\theta = \ell_p''(\hat\theta)\inv Q'(\hat\theta) + R
\end{equation} 
where the remainder $R$ is of smaller order in probability than the leading
term  under the assumption of uniform local continuity of $Q''$.
Therefore $\tilde\theta$ and $\hat\theta$ differ by $O_p(n\inv)$.
 
It is common practice to obtain standard errors for $\hat\theta$ via an
approximation of its covariance matrix with the inverse of the observed
information matrix $-\ell''(\hat\theta)\inv=I_O(\hat\theta)\inv$, say.
Combining the fact $\tilde\theta-\hat\theta=O_p(n\inv)$ with local continuity
of $Q''(\theta)$, we obtain the matching approximation
\begin{equation}   \label{e:var(MPLE)}
   \var{\tilde\theta} \approx -\ell''_p(\tilde\theta)\inv \,.
\end{equation}

%------------
\subsection{On the choice of $Q$}
 
The above formulation leaves an extremely wide set of options as for choice of
the penalty function. One way for selecting $Q$, or nearly equivalently for
selecting $M(\theta)= - Q'(\theta)$, is to require that the first order term
of the bias is eliminated. This is the route taken \citet{firth:1993} where
the requirement of bias reduction is adopted at the onset of the construction;
see also further work by \citet{kosm:firt:2009}.  If we insert $\pm\theta$ on
the left-hand side of \ref{e:tilde-hat} and compute expected values, then the
leading terms are
\[  
   \mathrm{bias}(\tilde\theta) \approx \mathrm{bias}(\hat\theta) 
     + I_E(\theta)\inv \:\E{M(\theta)}
\] 
where $I_E(\theta)=n\, i_E(\theta)$ is the expected information matrix and of
course computation of the expected value of $M(\theta)$ is void when $Q$ does
not depend on the data.  On equating the left side of this expression to $0$,
we obtain the condition
\[  
  \E{M(\theta)} =  -I_E(\theta)\,\mathrm{bias}(\hat\theta) 
\]
which must be completed by substitution of $\mathrm{bias}(\hat\theta)$ with
the first-order term of the MLE bias, given by \citet{cox:snel:1968}.  When
$M(\theta)$ does not depend on the data, we arrive at an estimating equation
of type \ref{e:SF-eqn}.

% Firth (1993, p.36): ``It is not an assumption of this work that bias 
%        reduction is always desirable.''

One difficulty with the bias reduction criterion for selecting $M$ is the
technical difficulty of working out the explicit expression of
$\mathrm{bias}(\hat\theta)$. In the skew-normal case, only the one-parameter
case leads to the relatively simple form \ref{e:M,a.p}, where however the
coefficients $a_p(\alpha)$ do not have an explicit expression.

Moreover, as  reminded by \citet{kosm:firt:2009}, ``Point
estimation and unbiasedness are, of course, not strong statistical principles.
The notion of bias, in particular, relates to a specific parameterization of a
model: for example, the unbiasedness of the familiar sample variance $S^2$ as
an estimator of $\sigma^2$ does not deliver an unbiased estimator of $\sigma$
itself''. We agree with this view, and in the development to follow the
requirement of unbiasedness will be taken into account but not in a
prescriptive form.

We conclude this section with a qualitatively motivated choice for $Q$ in the
case of a multivariate skew-normal distribution. It has repeatedly emerged
that many salient features of the family \ref{e:msn-pdf} depend on the
parameters via only the scalar quantity
\[
     \alpha_*^2= \alpha\T\bar\Omega\alpha
\]
where $\bar\Omega=\omega\inv\,\Omega\,\omega\inv$ is the correlation matrix
associated to $\Omega$.  The prominent role of $\alpha_*^2$ appears in a
number of results of \citet{azza:capi:1999} and of \citet{arel:azza:2008}; in
the latter paper the dependence is expressed indirectly via the monotonically
related quantity $ \beta_0^2=2\,\alpha_*^2/\{\pi+(\pi-2)\alpha_*^2\}$.

It is then natural to introduce a function $Q$ in \ref{e:logL.p} which depends
on $\theta$ only via $\alpha_*^2$. Combining this choice with the requirements 
\ref{e:Q-assume} and the consideration that a logarithmic form of dependence 
would keep the modification of the original log-likelihood to a minimum also
for diverging $\alpha_*$, we arrive at the formulation
\begin{equation} \label{e:Q}
     Q = c_1\,\log(1+ c_2\,\alpha_*^2) 
\end{equation}  
where $c_1$ and $c_2$ are positive constants. This is not yet a fully
specified penalty function, but the set on alternative options is now 
greatly reduced.

%------------
\subsection{On the choice of $Q$ in the skew-normal case}
\label{s:Q-SN}

We focus initially on the scalar skew-normal distribution.
In this case $Q$ and its first derivative take the form
\begin{equation} \label{e:Q-SN1}
   Q(\alpha) = c_1\,\log(1+c_2\alpha^2)\,,\qquad
   Q'(\alpha) =  2\,c_1\,c_2\frac{\alpha}{1+c_2\alpha^2}  \,.
\end{equation}  

Note that approximation \ref{e:M-BB} of $M(\alpha)$ is of type $-Q'(\alpha)$ 
with $c_1=3\,\pi^2/32$, $c_2=8/\pi^2$, but the intended use of $Q(\alpha)$ 
is not only for the one-parameter case to which \ref{e:M-BB} applies.

We want to develop an alternative approximation to $M(\alpha)$ defined by
\ref{e:M,a.p}. The reason for this search is partly to obtain an approximation
with stronger theoretical support and, more importantly, to explore a
direction which can extended to the skew-$t$ case which will be considered
later.

\begin{figure}
\centerline{
   \includegraphics[width=0.48\hsize]{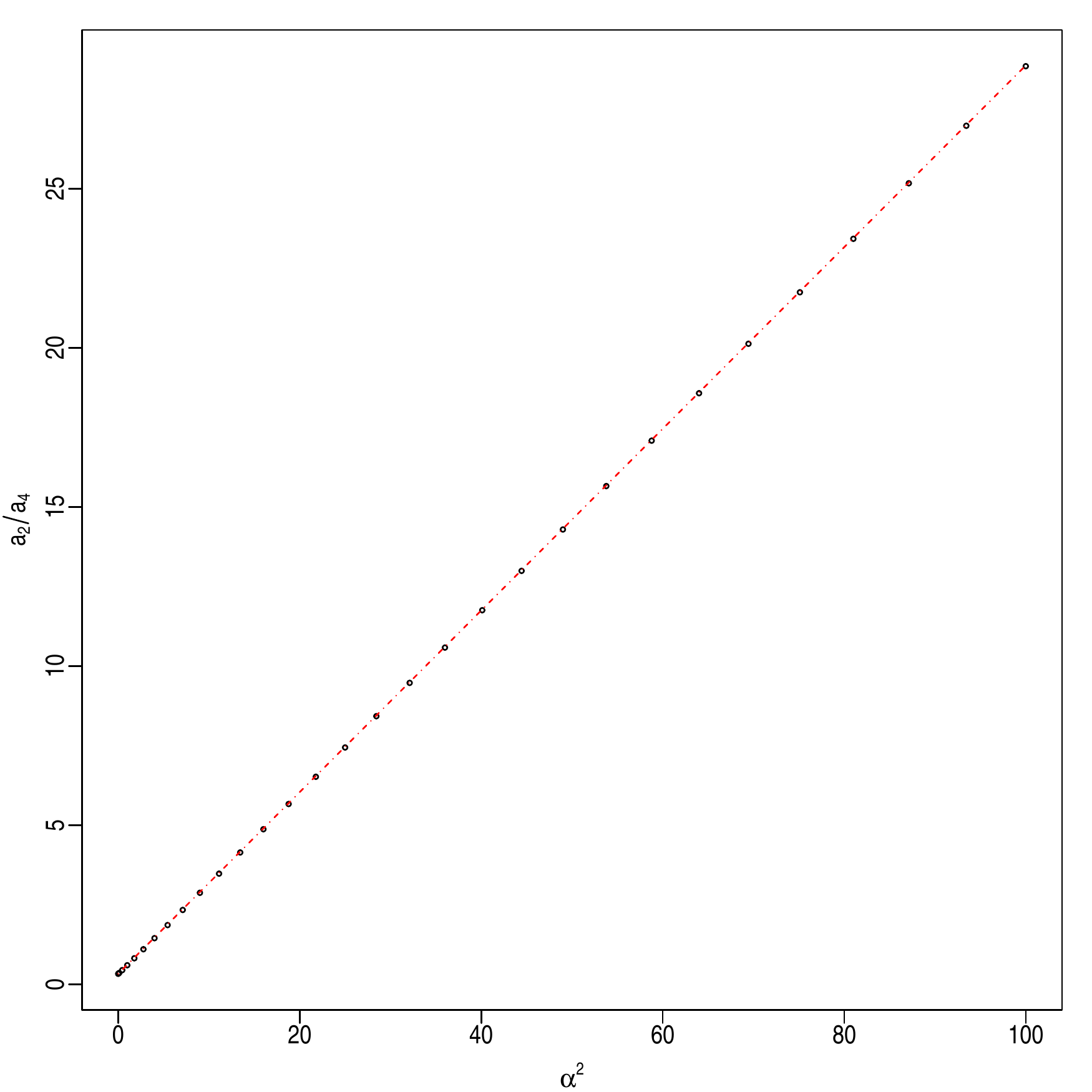}
   \hfill
   \includegraphics[width=0.48\hsize]{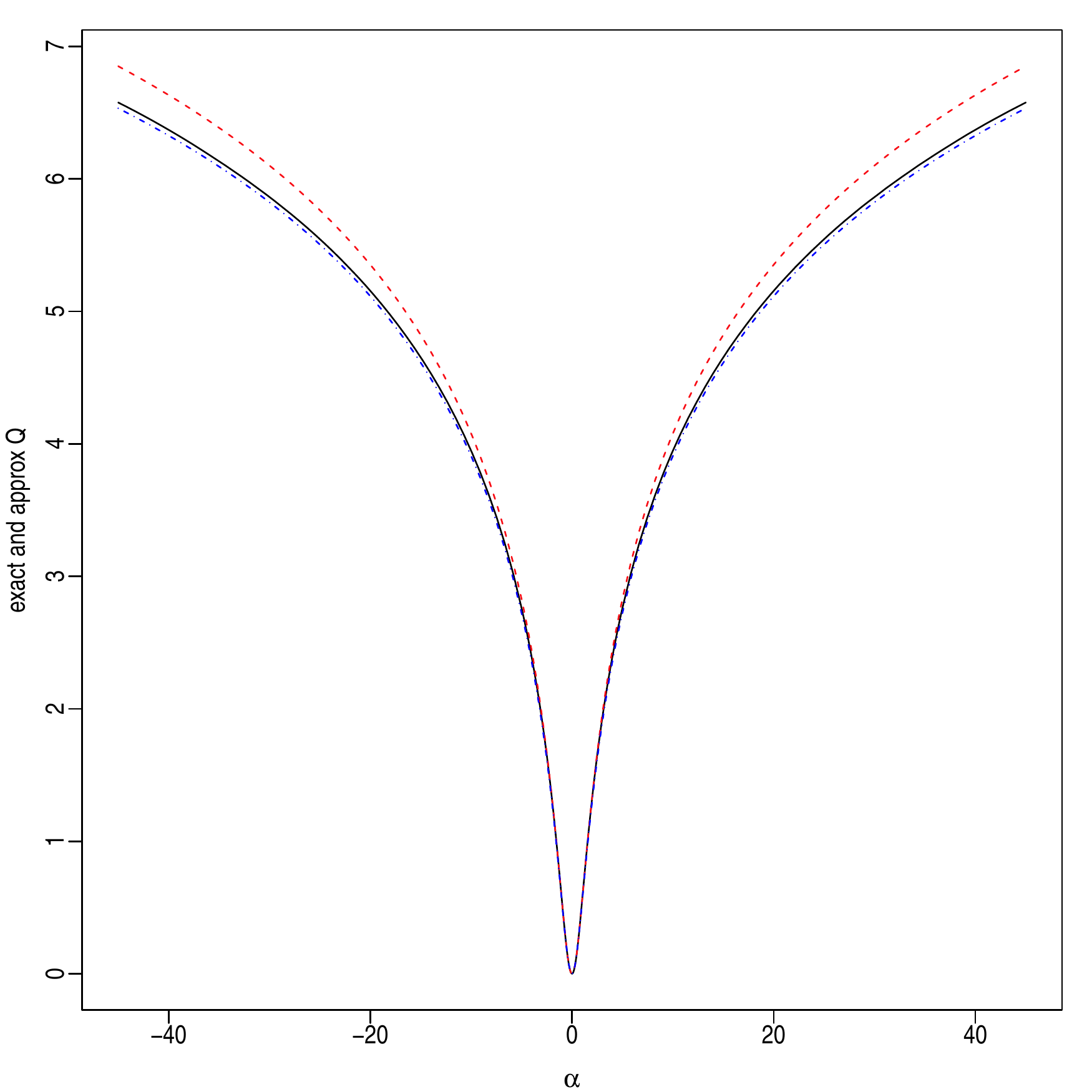}
   }
   \caption{\sl Left panel: values of $a_2(\alpha)/a_4(\alpha)$ for SN
     distributions, numerically evaluated at a grid of points, plotted versus
     $\alpha^2$ and superimposed approximating line.  Right panel: $Q$
     function obtained by numerical integration of $-M(\alpha)$ (continuous
     line), by integration of its approximation \ref{e:M-BB} (dashed line) and
     by $Q$  described in the text (dot-dashed line) }
\label{f:a2/a4&Q-SN}
\end{figure}

First, note that $a_2(\alpha)$ and $a_4(\alpha)$ are even functions of
$\alpha$.  This fact has been proved for $a_2(\alpha)$ by
\citet{lise:lope:2006} but the proof extends immediately to any $a_p(\alpha)$
with even $p$.  Hence $a_2/a_4$ depends on $\alpha$ only via $\alpha^2$.  Next
we observe that the numerical behaviour of $a_2/a_4$
% =\alpha/\{2\,M(\alpha)\}$
is remarkably linear with respect to $\alpha^2$ as shown by the left panel 
of Figure~\ref{f:a2/a4&Q-SN} which displays the value of $a_2/a_4$ 
at 31 equally spaced points of $\alpha$ between 0 and 10; 
the interpolating line will be described shortly. 

Therefore we  approximate $a_2/a_4$ by a function of the form 
$e_1 + e_2\alpha^2$ and we select $e_1$ and $e_2$ by matching
$a_2/a_4$ and  $e_1 + e_2\alpha^2$ at $\alpha^2=0$ and 
$\alpha^2\to\infty$. 
To this end, re-write $a_p(\alpha)$  as 
\[ 
  a_p(\alpha) = \sqrt{\frac{2}{\pi}}\: \frac{1}{(1+\alpha^2)^{p+1/2}}\:
     \E{X^p\,\zeta_1(\delta\,X)} 
\]
where $X\sim\N(0,1)$ and $\delta=\delta(\alpha)=\alpha/\sqrt{1+\alpha^2}$. 
Hence 
\[  \frac{a_2(\alpha)}{a_4(\alpha)} 
     = (1+\alpha^2) \frac{\E{X^2\,\zeta_1(\delta\,X)}}%
       {\E{X^4\,\zeta_1(\delta\,X)}}
     \approx e_1 + e_2 \,\alpha^2
\]
leading to       
\begin{eqnarray}
  e_1 &=& \frac{a_2(0)}{a_4(0)} 
       = \frac{\E{X^2 }}{\E{X^4 }}
       = \frac{1}{3} \, , \nonumber  \\
  e_2 &=& \lim_{\alpha^2\to\infty} \left\{ \frac{1+\alpha^2}{\alpha^2}\,
            \frac{\E{X^2\,\zeta_1(\delta\,X)}}{\E{X^4\,\zeta_1(\delta\,X)}}
               - \frac{e_1}{\alpha^2}\right\}
       = \frac{\E{X^2\,\zeta_1(X)}}{\E{X^4\,\zeta_1(X)}} 
       \approx 0.2854166                              \label{e:e1e2-SN}
\end{eqnarray}
where the final coefficient was obtained by numerical integration.  The line
plotted in the left panel of Figure~\ref{f:a2/a4&Q-SN} has intercept $e_1$ and
slope $e_2$.  
 
The right panel of Figure~\ref{f:a2/a4&Q-SN} displays three curves: the
continuous line is the curve obtained by numerical integration of $-M(\alpha)$
defined by \ref{e:M,a.p}, and it coincides up to the change of sign with the
continuous curve in Figure~1(b) of \citet{sartori:2006}; the dashed line is
the $Q$ function in \ref{e:Q-SN1} with $c_1=3\,\pi^2/32$, $c_2=8/\pi^2$,
corresponding to the integral of approximation \ref{e:M-BB}; the dot-dashed
line is the curve $Q$ with coefficients 
\begin{equation} \label{e:c1c2-SN}
   c_1=1/(4\,e_2)\approx 0.875913,\qquad c_2=e_2/e_1\approx 0.856250\,,
\end{equation}   
whose graph is barely distinguishable from the essentially exact continuous 
curve.

This choice of $Q$ with coefficients \ref{e:c1c2-SN} is motivated by the
Sartori-Firth formulation for the case $\SN(0,1,\alpha)$. However, we adopt
the same penalty function more generally, to the three-parameter case
$\SN(\xi,\omega^2,\alpha)$, since the motivation for introducing a
penalization of the log-likelihood function came solely from its behaviour
with respect to $\alpha$. When this procedure is applied to the frontier data,
the estimates of $(\xi,\omega,\alpha)$ are $(-0.034, 1.165, 6.256)$, quite
close to the true parameters $(0,1,5)$ and also close to the Sartori 
values $(-0.106, 1.234, 6.243)$, whose first two components coincide with
the MLE.
% if one bears in mind that the sample  size $n=50$ must be regarded 
% as small in the context of  distributions
% effectively depending on moments up to the third order. 
The  graphical outcome of the MPLE is represented by 
the dot-dashed curve in the left panel of Figure~\ref{f:frontier-data}. 
It is also worth mentioning that a shape parameter $\alpha=6.256$ 
corresponds to an index of skewness $\gamma_1=0.899$,
very close to the sample index of skewness, $0.902$.

% In addition, as noted by \citet{sartori:2006}, the MLE's of $\xi$ 
% and $\omega$ are satisfactory estimates of the parameters, even when
% $\hat\alpha$ diverges, and a penalization of the log-likelihood to adjust
% these components seems superfluous also from the numerical perspective.  The
% fact that $\hat\xi$ and $\hat\omega$ are hardly affected when $\hat\alpha$
% varies is illustrated, in one specific instance, by 
% Figure~\ref{f:frontier-mle}.

Consider now a $d$-dimensional skew-normal distribution \ref{e:msn-pdf},
initially  in  the case of a  location- and scale-free variable 
$Z\sim\SN_d(0, \bar\Omega,\alpha)$ where $\bar\Omega$ is a correlation matrix. 
Recall the canonical transformation $Z^*= A^*\,Z$ introduced in Proposition~4
of \citet{azza:capi:1999}, such that $Z^*$ has $d$ independent
components of which one (the first one, say) has distribution
\[  
    \SN(0, 1, \alpha_*), \qquad 
    \alpha_* = \left(\alpha\T\bar\Omega \alpha\right)^{1/2}\,, 
\]
and the other $d-1$ components are $\N(0,1)$.  More specifically, an
explicit expression of the transformation, given in the proof available in
the full version of the paper, is
\begin{equation} \label{f:Z-canonical}
     Z^* = (C\inv P)\T \:Z  % (C\inv P)\T \omega\inv(Y-\xi) =
\end{equation}
where $C$ is such that
\[ \bar\Omega= C\:C\T \] and $P$ is an orthogonal matrix whose first column is
proportional to $C\alpha$.  We can then write
\[  
    Z^* \sim \SN_d(0, I_d, \alpha_{Z^*}), \qquad
   \alpha_{Z^*}=(\alpha_*, 0, \dots,0)\T \,.
\]
Assume now that a random sample $z= (z_1,\dots,z_n)$ is drawn from 
$Z\sim\SN_d(0,\bar\Omega, \alpha)$. To estimate its parameters, we can 
proceed as follows.
\begin{itemize}
\item The sample $z$ can be converted into an equivalent sample
  $z^*=(z_1^*,\dots,z_n^*)$ drawn from $Z^*\sim\SN_d(0, I_d, \alpha_{Z^*})$ on
  setting
\[ z_i^* = (C\inv P)\T \:z_i \qquad (i=1,\dots,n).\] The determinant of the
Jacobian is $ \det(C\T \:P) = \det(C)= \det(\bar\Omega)^{1/2} \,. $
\item We now have a sample of size $n$ from $\SN(0,1,\alpha_*)$ and $d-1$
  samples of size $n$ from $\N(0,1)$. For the first sample, we adopt the
  above-described scheme, hence the log-likelihood function is as on
  \ref{e:logL-sn1} with $\alpha$ replaced by $\alpha_*$ and 
  the penalty function is \ref{e:Q} with coefficients which can be 
  taken as in \ref{e:c1c2-SN}.
\item Now we can revert back the $z^*$ sample to the original $z$, for which
  we write the usual log-likelihood, except that in this process we have
  introduced the penalty factor for $\alpha$. In conclusion the penalized
  log-likelihood is
\begin{eqnarray}
  \ell_p(\theta)  
     &=&  \ell(\theta) - c_1\,\log(1+c_2\alpha_*^2)    \nonumber  \\
     &=&  \sum_{i=1}^n \left(\log\phi_d(z_i;\bar\Omega)
          + \zeta_0(\alpha\T z_i) \right) - c_1\,\log(1+c_2\alpha_*^2) \,.
        \label{e:logL.p-SN0}
\end{eqnarray}
\end{itemize}
The following remarks apply.  First, the transformations from $z$ to $z^*$ and
then back to $z$ are conceptual steps which serve only as an argument to
introduce the penalty factor, and support its present form, and they do not
need to be actually performed.  Second, note that, although $d$ does not
appear explicitly in the penalty factor, it does have an effect since
$\alpha_*^2$ reflects $d$ indirectly, via the increase of the number of
summands in its expression.  As a simple example, take the case where $\alpha$
is a vector with $d$ identical components $\alpha_0$ and $\bar\Omega=I_d$,
then $\alpha_*^2 = d\,\alpha_0^2$.
  
Now move to the general case of \ref{e:msn-pdf} and consider a random sample
$y=(y_1,\dots,y_n)$ from $Y\sim \SN_d(\xi, \Omega, \alpha)$.  By adapting the
$\ell(\theta)$ term in \ref{e:logL.p-SN0} for the presence of location and
scale parameters, we arrive at
\begin{eqnarray} 
 \ell_p(\theta) 
     % &=&  \ell(\theta) - c_1\,\log(1+c_2\alpha_*^2) \label{e:logL.p-SN} \\
     &=&  \sum_{i=1}^n \left[\log\phi_d(y_i-\xi_i;\Omega)
          + \zeta_0(\alpha\T \omega\inv(y_i-\xi_i)) \right]
           - c_1\,\log(1+c_2\alpha_*^2)                \label{e:logL.p-SN}
\end{eqnarray}
where $\theta\T=(\xi\T,(\vech\Omega)\T,\alpha\T)$; here $\vech$ is the operator
which stacks the lower triangle of a matrix in a vector.

%------------
\subsection{LRT-type statistic and another estimate}
\label{s:LRT+bar.theta} 

Consider now the likelihood-ratio test (LRT) statistics in its standard
version and the analogous one for the penalized log-likelihood \ref{e:logL.p},
that is
\[ 
    W  = W(\theta)  = 2\{\ell(\hat\theta)-\ell(\theta)\},\qquad  
    W_p = W_p(\theta)= 2\{\ell_p(\tilde\theta)-\ell_p(\theta)\} \,.
\]    
From the results of  Section~\ref{s:logL.p}, we can say that the null 
distribution of $W_p$ is of $\chi^2$ type, similarly to $W$.

\begin{figure}
\centerline{ 
   \includegraphics[width=0.48\hsize]{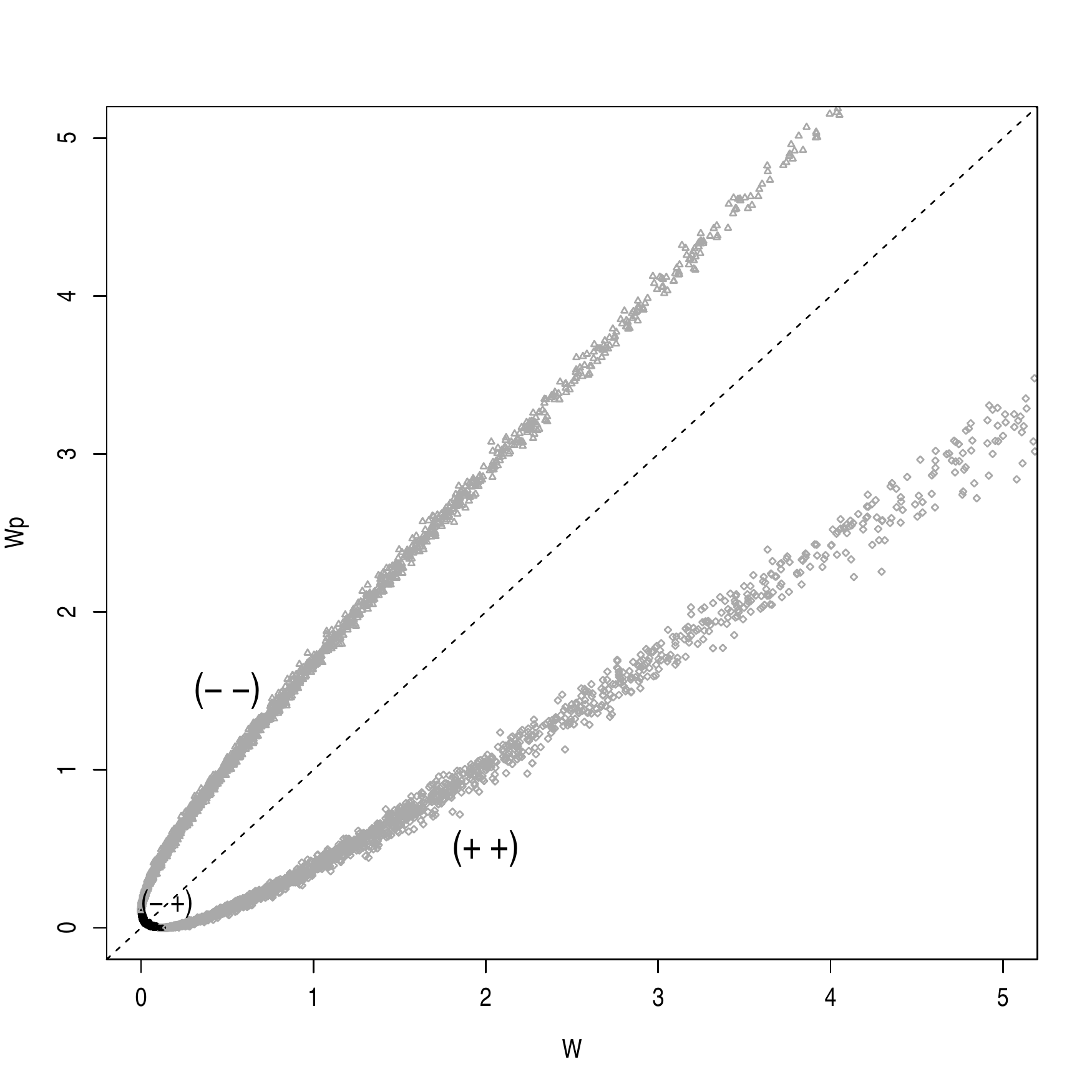}
   \hfill
   \includegraphics[width=0.48\hsize]{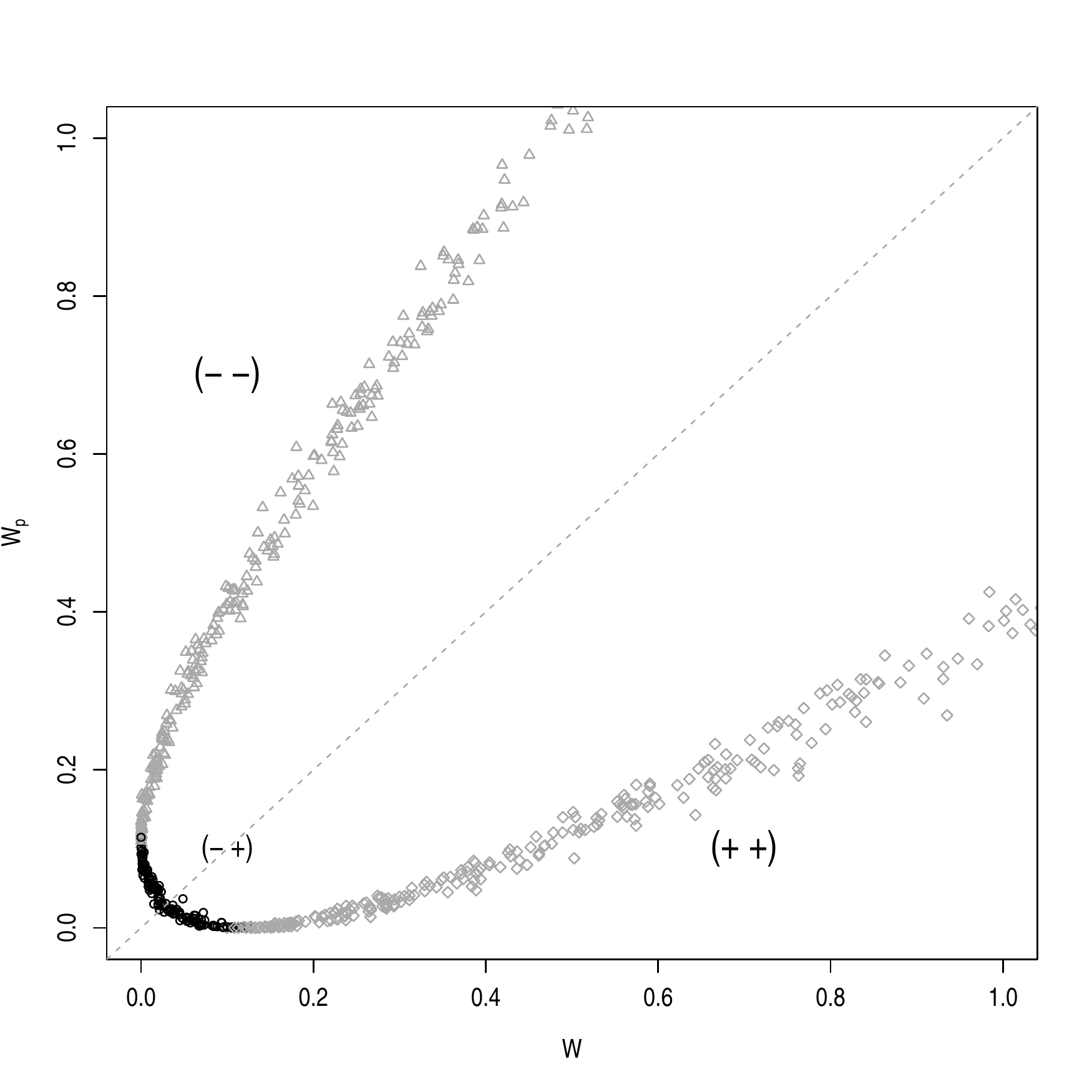}
   }
   \caption{\sl Scatter plot of simulated values of $W(\alpha)$ and
     $W_p(\alpha)$ for samples of size $n=100$ from $\SN(0,1,\alpha)$ when
     $\alpha=3$.  Samples where both $\hat\theta$ and $\tilde\theta$
     overestimate $\alpha$ are marked by grey diamonds, those where both
     underestimate are denoted by grey triangles, those with mixed signs are
     denoted by black circles.  The right-side plot refers a subset of the
     sampled values and it shows the enlarged picture of a smaller area. The
     dashed line is the identity.  }
\label{f:WpW-scatter}
\end{figure}

Both intuition and the results of Section~\ref{s:logL.p} suggest that $W$ and
$W_p$ must be strongly associated. Some numerical exploration confirms this
idea but it also exhibits that the type of dependence is quite peculiar. This
is illustrated in the left plot of Figure~\ref{f:WpW-scatter} which refers to
a set of 5000 samples of size $n=1000$ from the distribution $\SN(0,1,\alpha)$
with $\alpha=3$; hence in this case $\theta$ is $\alpha$.  To increase
readability, the right plot of the same figure displays a subset of the
earlier points over a reduced plotting area. The obvious feature of this
figure is that the joint distribution of $(W, W_p)$ is strongly concentrated
along two branches.  A closer inspection indicates that the top branch is made
of points where both $\hat\alpha$ and $\tilde\alpha$ underestimate the true
$\alpha=3$, the bottom branch is composed by points where both $\hat\alpha$
and $\tilde\alpha$ overestimate, and the darker points around in the bottom
left corner are those where $\hat\alpha-\alpha$ and $\tilde\alpha-\alpha$ take
opposite signs.  Note that the points with opposite signs of the estimation
error are whose with smaller values of both $W$ and $W_p$, hence those where
the estimation error is smaller is size.

The pattern displayed in Figure~\ref{f:WpW-scatter} appears also in other
simulation experiments.  This behaviour, combined with the final remark of the
previous paragraph, suggests an alternative estimate $\bar\theta$ defined as a
solution of $W=W_p$, written more explicitly as
\begin{equation}  \label{e:Wp=W}
  \bar\theta= \{\theta :  W(\theta)- W_p(\theta) =0 \}   
\end{equation}
or equivalently
\begin{equation}   \label{e:Q=q(y)}
   \bar\theta= \{\theta :  Q(\theta)= q(y)\}, \qquad
   q(y) = \ell(\hat\theta) -\ell(\tilde\theta) + Q(\tilde\theta)  \,.
\end{equation}
Note that $q(y)\ge0$, with strict inequality if we exclude limiting cases.
For the existence of this solution we need that (i)~$W_p$ and $W$ are finite,
and (ii)~there exist points of the parameter space with opposite signs of
$W_p-W$. In the context of skew-normal distribution, $\ell(\hat\theta)$ 
and $\ell_p(\tilde\theta)$ are bounded, so condition (i) holds. As for
condition (ii), it holds because of the following facts:
\begin{eqnarray}
  W(\tilde\theta)>0,  &&    W_p(\hat\theta)>0,   \nonumber \\
  W_p(\tilde\theta) = 0, &&   W(\hat\theta)=0, \nonumber \\
  % \Longrightarrow 
  W_p(\tilde\theta)-W(\tilde\theta) < 0  , && % \qquad 
  W_p(\hat\theta)-W(\hat\theta) >0  \,. \label{e:WpW-signs}
\end{eqnarray}

In the multiparameter case,  \ref{e:Wp=W} defines a surface in the
parameter space, not a single point. In this case we complement \ref{e:Wp=W}
with the condition that $\bar\theta$ must lie on the segment joining
$\hat\theta$ and $\tilde\theta$. Since inequalities \ref{e:WpW-signs}
ensure that $\hat\theta$ and $\tilde\theta$ lie on opposite sides of 
the surface, this intersection point exists. From the computational viewpoint,
$\bar\theta$ can be located  efficiently, via a one-dimensional search
along this segment, irrespectively of the dimension of $\theta$.

When $Q$ is chosen of the form \ref{e:Q}, \ref{e:Q=q(y)} takes a
simple form, since the solution of $c_1\log(1+c_2\alpha_*^2)=q(y)$ corresponds
to the equation of an ellipsoid, that is  $\alpha\T\bar\Omega\alpha = r(y)$,
where $r(y)=[e^{q(y)/c_1}-1]/c_2$.

\subsection{Simulation study} 

The above estimation methods have been studied via numerical simulations.  The
first study has considered samples from $\SN(0,1,\alpha)$ where $\alpha$ was
the only parameter to be estimated, and the true value was $\alpha=5$. For
each generated sample, four estimators have been computed: classical MLE, MPLE
with penalty \ref{e:Q-SN1} and coefficients \ref{e:c1c2-SN}, the
Sartori--Firth estimator defined by \ref{e:SF-eqn}-\ref{e:M,a.p}, the
estimator defined by the condition $W_p=W$ in \ref{e:Wp=W}. These estimates
have been computer for $10^6$ replicated samples, for each of sample sizes
$n=50, 100, 250, 350, 500, 1000$.

\begin{figure}
\centerline{
  \includegraphics[width=0.49\hsize]{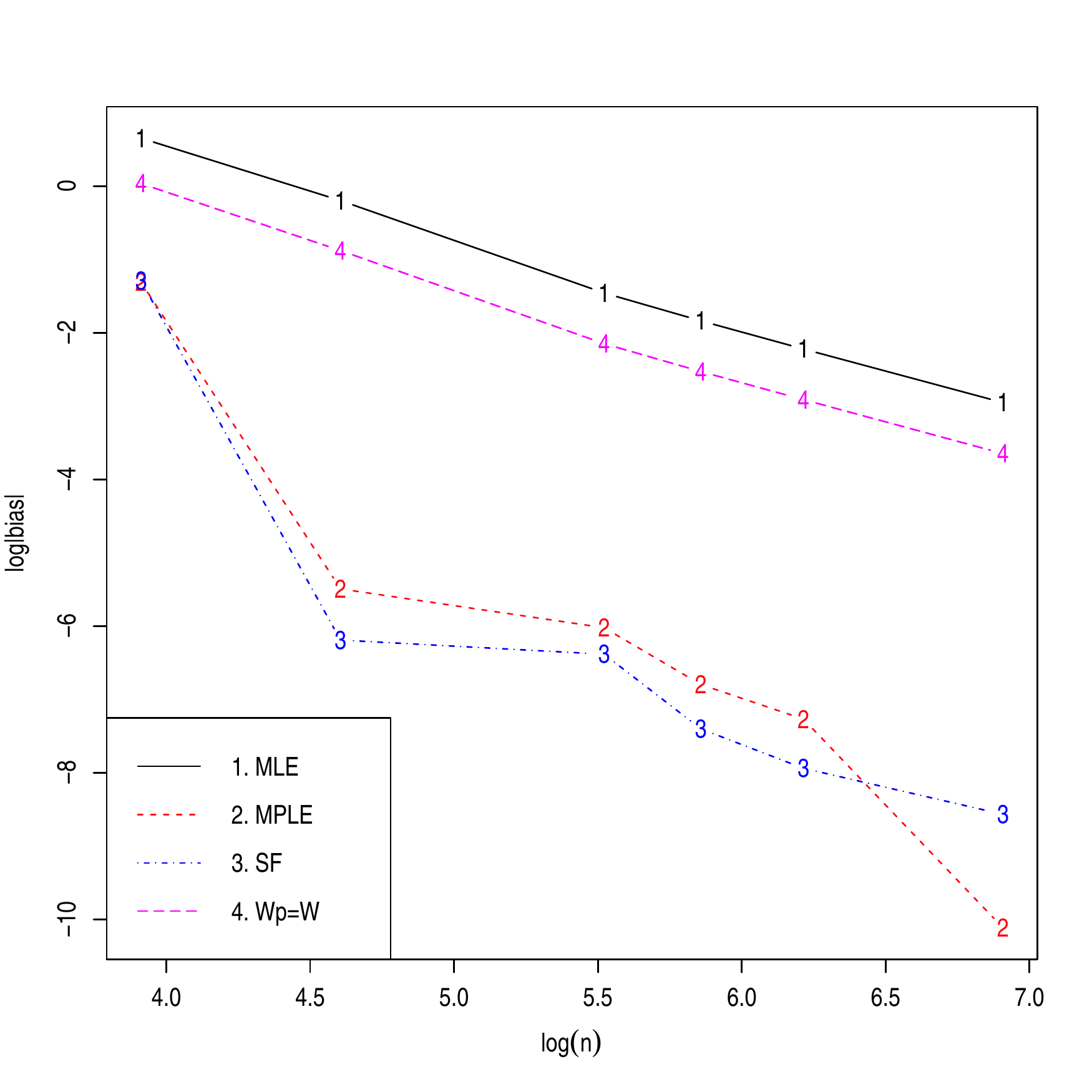}\hfill
  \includegraphics[width=0.49\hsize]{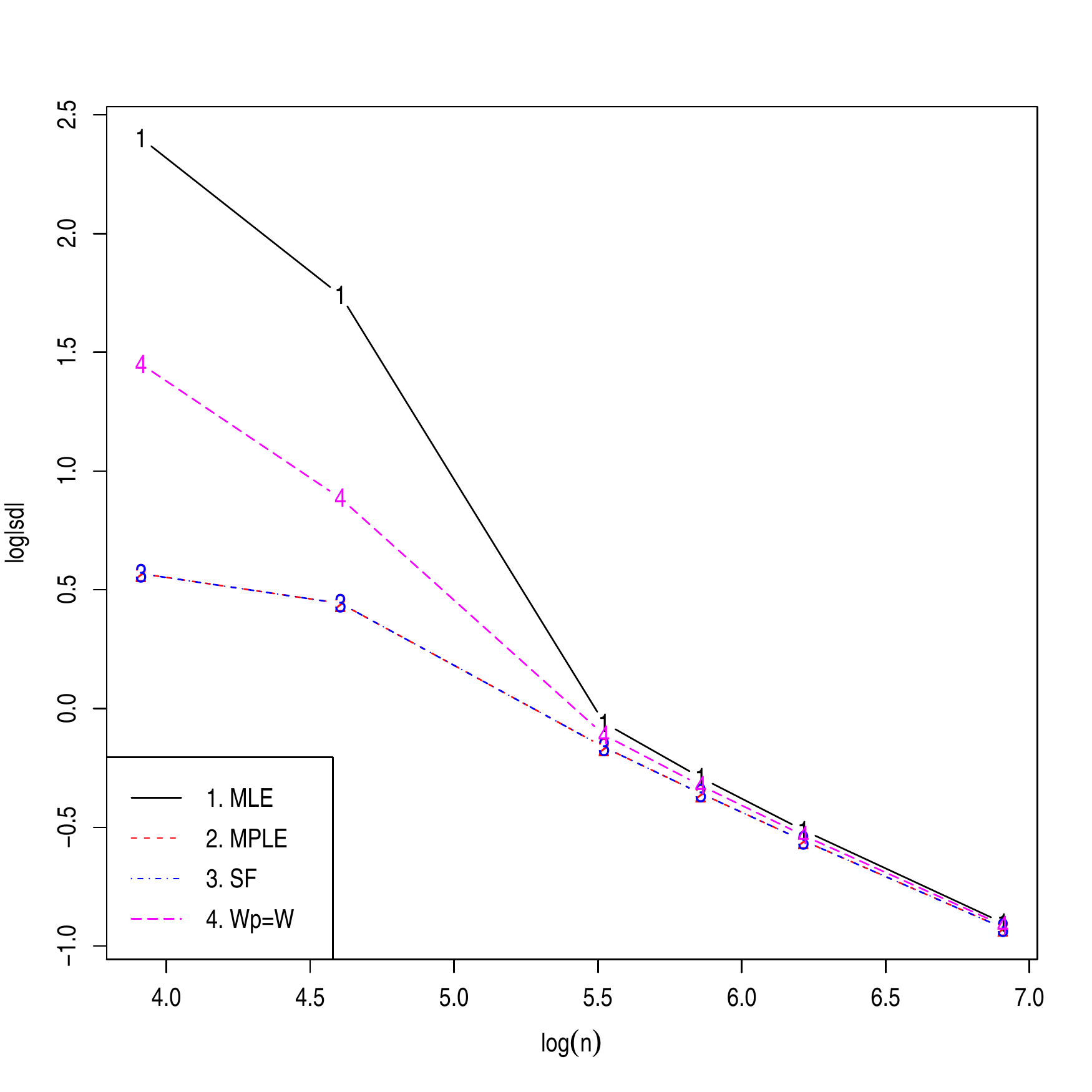}}
\centerline{
  \includegraphics[width=0.49\hsize]{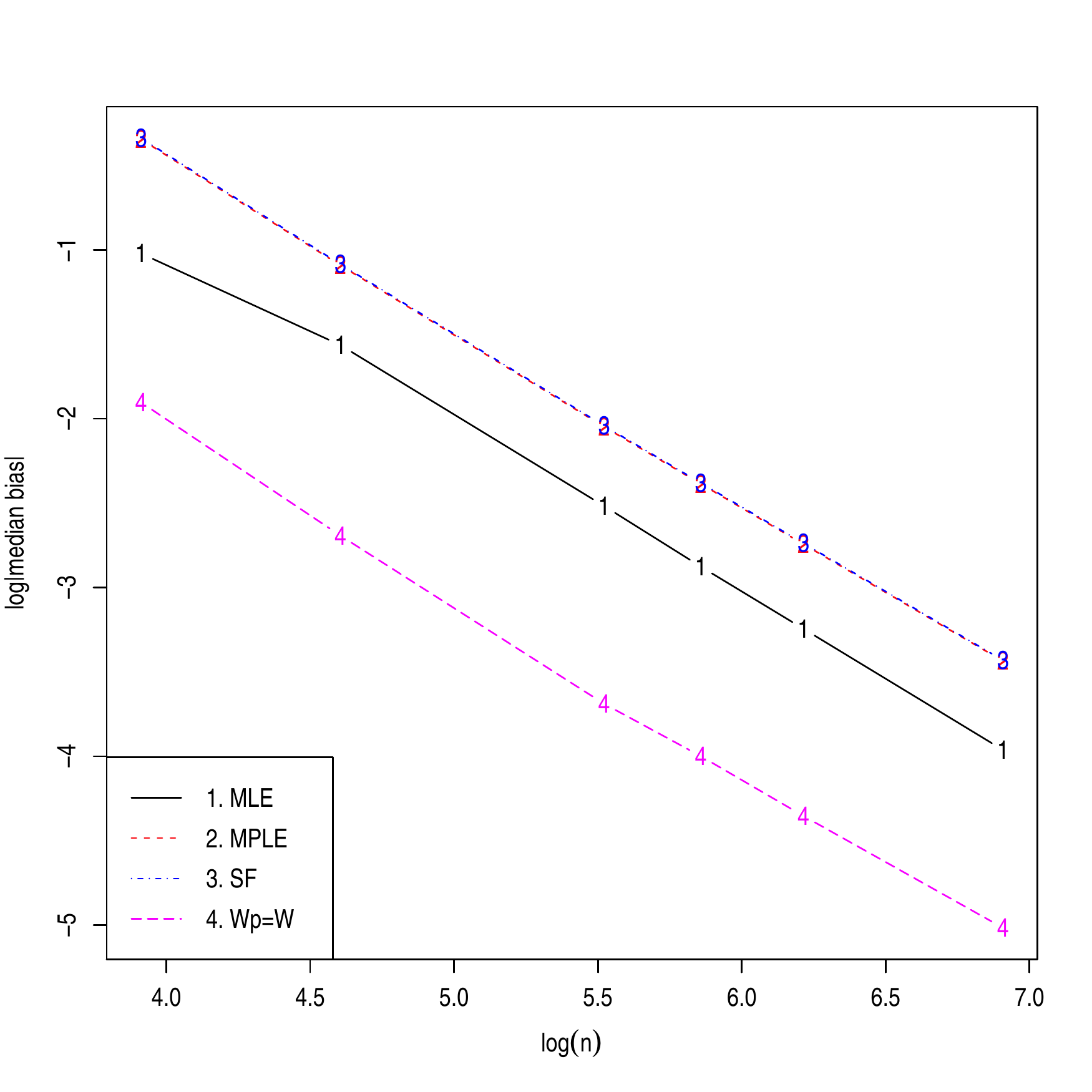}\hfil
  \includegraphics[width=0.49\hsize]{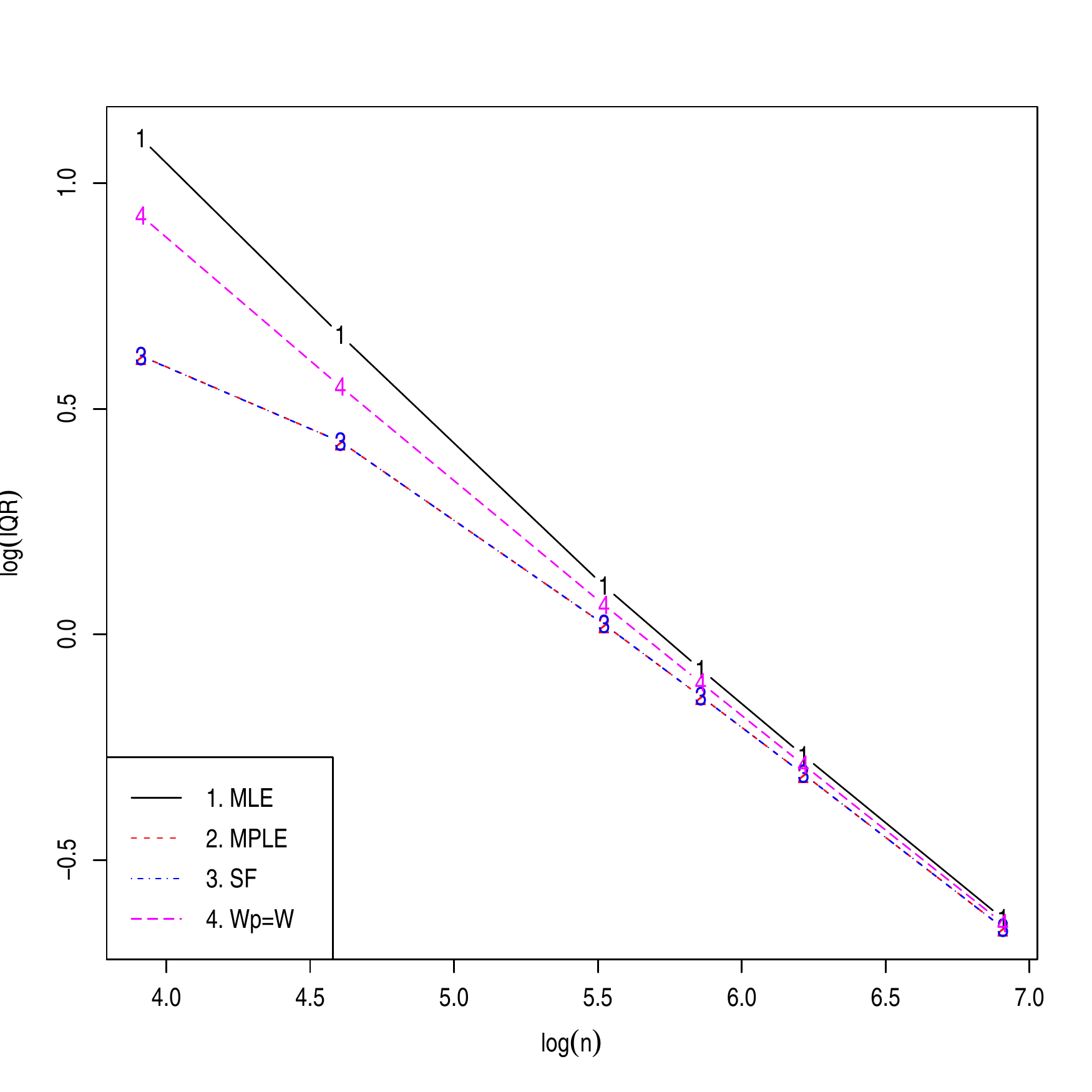}}
\caption{\sl Simulation study of the distributions of four estimators for
  samples of size $n=50, 100, 250, 350, 500, 1000$ from $\SN(0,1,\alpha)$ when
  $\alpha$ only is estimated and the true value is $\alpha=5$; for each sample
  size $10^6$ replicates have been generated.  The four estimators are 1:MLE,
  2:MPLE, 3:SF, 4:$W_p=W$, but on three of the four panels MPLE and SF are
  graphically coincident.  Top left panel: $\log|\mathrm{bias}|$, top right:
  log standard deviation, bottom left: $\log|\mathrm{median~bias}|$, bottom
  right: log IQR; in all case the horizontal axis represents $\log n$.  }
\label{f:simul-SN1-alpha}
\end{figure}

The final outcome is summarized graphically in the four panels of
Figure~\ref{f:simul-SN1-alpha}, where the curves associated to the estimators
are numbered $1$ to $4$.  Four log-transformed summary quantities are plotted
versus $\log n$; they are: absolute bias (top left), standard deviation
(top right), absolute median bias (bottom left), inter-quartile range
(bottom right).  The cases where $\hat\alpha$ diverged have been excluded from
the computation of these summaries, in agreement with \citet{sartori:2006} and
\citet{baye:bran:2007}. 

The doubly logarithmic scale has been adopted for simplifying interpretation
of the curves. This is especially so for the top left panel, since the bias is
expected to decrease at rate $n\inv$ for the MLE, and at rate $n^{-2}$ for the
Sartori-Firth method and its close approximation MPLE.  We do in fact observe
that the slopes of these curves are close to $-1$ and $-2$ respectively.  The
top left panel confirms the theoretical expectations, displaying a clear
improvement of SF and MPLE over MLE; the estimator \ref{e:Wp=W} has a bias
somewhat lower than MLE but decreasing at the same rate. MPLE and SF are very
similar to each other, with only some discrepancy at the right end, with
$n=1000$, where the bias is very small indeed and the numerical approximation
involved in evaluating the coefficients $a_p(\alpha)$ in \ref{e:M,a.p} may
have perturbed slightly the exact implementation of the SF method.

The message emerging from the bottom left panel, which plots the logarithm of
the median bias, is radically different. Estimator \ref{e:Wp=W} is markedly
preferable to the others, MLE is second best, and the other two are equivalent
up to the point that their curves are superimposed. For all the curves the
slope is near to $-1$, which means ad the rate of decrease $n\inv$ for the
median bias, and the differences are only in the intercepts. Median
unbiasedness is not often considered in theoretical work, presumably because
it is a more difficult aspect to evaluate compared to mean unbiasedness; it
has however the important advantage over mean unbiasedness that it is
preserved under monotone parameter transformation.

The two right-side panels convey very similar messages as for variability of
the four contenders: SF and MPLE have the smallest variability, both on the
scale of standard deviation and of the inter-quartile range, and they are
essentially equivalent to each other with superimposed curves; MLE has the
largest variability and estimator \ref{e:Wp=W} sits in between the others.
Differently from the left-side plots, however, here the differences vanish as
$n$ diverges, and they are effectively almost negligible from $n=250$ onwards.

The usual combination of bias and variance is given by the mean square error,
or by its square root. If the logarithm of either of them is plotted versus
$\log n$, the graphical appearance is virtually identical to the top right
panel of Figure~\ref{f:simul-SN1-alpha}.  If comparisons are based on
moment-based quantities, the overall indications are then that (i)~MPLE and SF
are preferable to the others, at least for small and moderate $n$, (ii)~MPLE
and SF are effectively equivalent.  It is less clear how to combine the two
bottom panels into a single summary plot, but we note that also here the
vertical scale of right-side panel has a larger order of magnitude than the
left-side panel; hence its behaviour would dominate in any reasonable
combination of the two.

In a second set of simulations, the data have still been sampled from a
skew-normal distribution but now all three parameters are regarded as unknown,
so that the reference set of distributions is of type $\SN(\xi, \omega^2,
\alpha)$.  To ease comparisons, the parameter values and the sample sizes have
been taken to be the same of Table~2 of \citet{sartori:2006}, that is $\xi=0$,
$\omega=1$, $\alpha=5, 10$, with sample sizes $n=50, 100, 200$, but here a
substantially larger number of replicates has been generated for each sample
size, namely $10^5$.  In all cases the location and scale parameters were
$\xi=0$ and $\omega=1$.  Table~\ref{t:simul-SN1-3param} reports the summary
values for three estimators, that is MLE, MPLE and \ref{e:Wp=W}; for the last
one, only the estimate of $\alpha$ is given, since the first two components of
the estimate coincide with those of MPLE.

Inspection of the Table~\ref{t:simul-SN1-3param} confirms the improvement of
$\tilde\alpha$ over $\hat\alpha$ and its overall similarity of the bias of
$\tilde\alpha$ with the analogous entry in Table~2 of \citet{sartori:2006}.
Notice that, $\hat\alpha$ and $\tilde\alpha$ are now no longer essentially
coincident as they were in the one-parameter case.  In interpreting the bias
and standard deviation of $\hat\alpha$ one must bear in mind that this have
been computed excluding the case with diverging estimates; in practical term,
we have taken $|\hat\alpha|>100$ as an indication of a diverging estimate.
The exclusion of the diverging estimates has a relevant effect especially in
the present setting where the probability of this event appears to be
appreciably higher than in the one-parameter case; for instance with $n=50$
and $\alpha=5$ this probability is now about $0.139$, while it is only $0.039$
in the one-parameter case examined earlier.  Bias and variability of
$\tilde\xi$ and $\tilde\omega$ are somewhat higher than those of the MLE's,
$\hat\xi$ and $\hat\omega$, at least for $n=50$ and to some extent for
$n=100$.  It is then conceivable to adopt $\hat\xi$ and $\hat\omega$ as
estimates of location and scale, and $\tilde\alpha$ for estimating shape,
similarly to the strategy of Sartori.  This choice is advantageous as for
formal properties, but it has the logical drawback of the lack of a unique
estimation criterion. Also in this setting, the estimate $\bar\alpha$ has low
median bias, but this is paid by a substantial increase in variability.

%-- http://blog.mikezhang.com/dcolumn/processtable.cgi
\begin{table}
\begin{center}
\caption{\sl Summary quantities of the distribution of MLE, MPLE and
\ref{e:Wp=W} estimating the parameters $(\xi, \omega, \alpha)$ of a
distribution $\SN(0,1,\alpha)$ when $\alpha=5, 10$, based on a sample of
size $n=50, 100, 200$.  All entries are based on $10^5$ replicated
samples. }
\label{t:simul-SN1-3param}
\vspace{0.5ex} \small
\begin{tabular}{crl*{7}{d}}	%%zxqdcolumn
%----
\hline
\vphantom{\rule[-1ex]{0ex}{3.5ex}}
% \multicolumn{1}{c}{
$\alpha$ & $n$ &  & 
   \multicolumn{1}{c}{\ensuremath{\hat\xi}}  &  
   \multicolumn{1}{c}{\ensuremath{\hat\omega}} & 
   \multicolumn{1}{c}{\ensuremath{\hat\alpha}} &
   \multicolumn{1}{c}{\ensuremath{\tilde\xi}} & 
   \multicolumn{1}{c}{\ensuremath{\tilde\omega}} & 
   \multicolumn{1}{c}{\ensuremath{\tilde\alpha}}  & 
   \multicolumn{1}{c}{\ensuremath{\bar\alpha}}
\\
\hline  %------------
\vphantom{\rule[-1ex]{0ex}{3ex}}
5 & 50 
& mean bias   & 0.0235& -0.0209& 1.472 & 0.1740& -0.1252& -1.411& 1.808 \\
&& median bias& 0.0028& -0.0160& 0.124 & 0.0884& -0.1023& -1.726&-0.610 \\
&& std.~dev.  & 0.1502& 0.1411 & 4.974 & 0.2661&  0.1677&  2.600& 7.641 \\
&& $\pr{\hat\alpha=\infty}$ & && 0.139\\
& 100 
& mean bias   & 0.0060&-0.0075& 1.442 & 0.0534& -0.0499& -0.456& 0.866 \\
&& median bias& 0.0002&-0.0066& 0.263&  0.0379 & -0.0456& -0.872 &-0.306 \\
&& std.~dev.  & 0.0839& 0.0959& 4.896& 0.1118& 0.1011& 2.320& 5.615 \\
&& $\pr{\hat\alpha=\infty}$ &&& 0.023\\
& 200 
& mean bias   & 0.0029 &-0.0036 & 0.592 & 0.0216&-0.0222&-0.195 & 0.197 \\
&& median bias& 0.0005 &-0.0035 & 0.150 & 0.0181 &-0.0216&-0.410 & -0.152 \\
&& std.~dev.  & 0.0544 & 0.0656 & 2.508&  0.0558 & 0.0655 & 1.540& 2.254 \\
&& $\pr{\hat\alpha=\infty}$ & & &  0.001\\
%------ 
10& 50 
& mean bias   & 0.0225 &-0.0235 & 0.788&  0.1249& -0.1049& -3.992& 3.497 \\
&& median bias& 0.0076& -0.0210& -1.001&  0.0685& -0.0898& -4.586& -1.070 \\
&& std.~dev.  & 0.0910& 0.1213& 6.874&  0.2038& 0.1461& 3.769& 11.775 \\
&& $\pr{\hat\alpha=\infty}$ & & &  0.345\\
& 100
& mean bias   & 0.0047&-0.0069& 3.274& 0.0393&-0.0418&-1.462& 3.591 \\
&& median bias&-0.0004&-0.0067& 0.527& 0.0299&-0.0404&-2.481&-0.638 \\
&& std.~dev.  & 0.0560& 0.0830& 9.399& 0.0677& 0.0845& 4.580& 12.840 \\
&& $\pr{\hat\alpha=\infty}$ & & &  0.129\\
& 200
& mean bias   & 0.0015&-0.0025& 2.575& 0.0170&-0.0190&-0.422& 1.583 \\
&& median bias&-0.0026&-0.0026& 0.563& 0.0140&-0.0188&-1.275&-0.360 \\
&& std.~dev.  & 0.0374& 0.0581& 8.072& 0.0375& 0.0574& 4.160 & 8.738\\
&& $\pr{\hat\alpha=\infty}$ & & &  0.021\\
\hline %-----------------
\end{tabular}
\end{center}    
\end{table}  

% ======================================================
\section{Extension to the skew-$t$ distribution} \label{s:ST}

A distribution closely related to the skew-normal  is the skew-$t$ whose
density in the scalar case is
\begin{equation} \label{e:dst}
    \frac{2}{\omega}\, t\left(\frac{x-\xi}{\omega}; \nu\right)\:
    T\left(\alpha\:\frac{x-\xi}{\omega}\:\sqrt{\frac{\nu+1}{\nu+Q_x}};
         \nu+1\right),
   \qquad x\in\Real,
\end{equation}  
where $t(\cdot;\nu)$ is the Student's $t$ density with $\nu>0$ degrees of
freedom, $T(\cdot;\nu+1)$ is the $t$ distribution function for $\nu+1$ degrees
of freedom and $Q_x=\omega^{-2}(x-\xi)^2$; here $\nu$ is a positive value
which can be non-integer. This distribution allows regulation of the tail
thickness via the additional parameter $\nu$. If a continuous random variable
$Y$ has density function \ref{e:dst}, we write $Y\sim\ST(\xi,\omega^2, \alpha,
\nu)$.

Initial work for applying the method of Firth to the case of a skew-$t$
distribution has been done by \citet{sartori:2006}, who has considered
estimation of $\alpha$ when the other parameters are known. For the
three-parameter case ($\xi, \omega, \alpha$) with known $\nu$, Sartori has
proposed a two-step procedure, similarly to the skew-normal case. This
direction has been explored further by \cite{lago:jime:2012} who have shown
that the corresponding estimating equation is of the same form $M(\alpha)=0$
as in \ref{e:M,a.p}, with $a_2$ and $a_4$ replaced by suitably modified
expressions which depend on $\nu$ besides $\alpha$.  Moreover, they prove that
$M(\alpha)=0$ always admits a finite solution when $\nu>2$.

\begin{figure}
\centerline{
   \includegraphics[width=0.48\hsize]{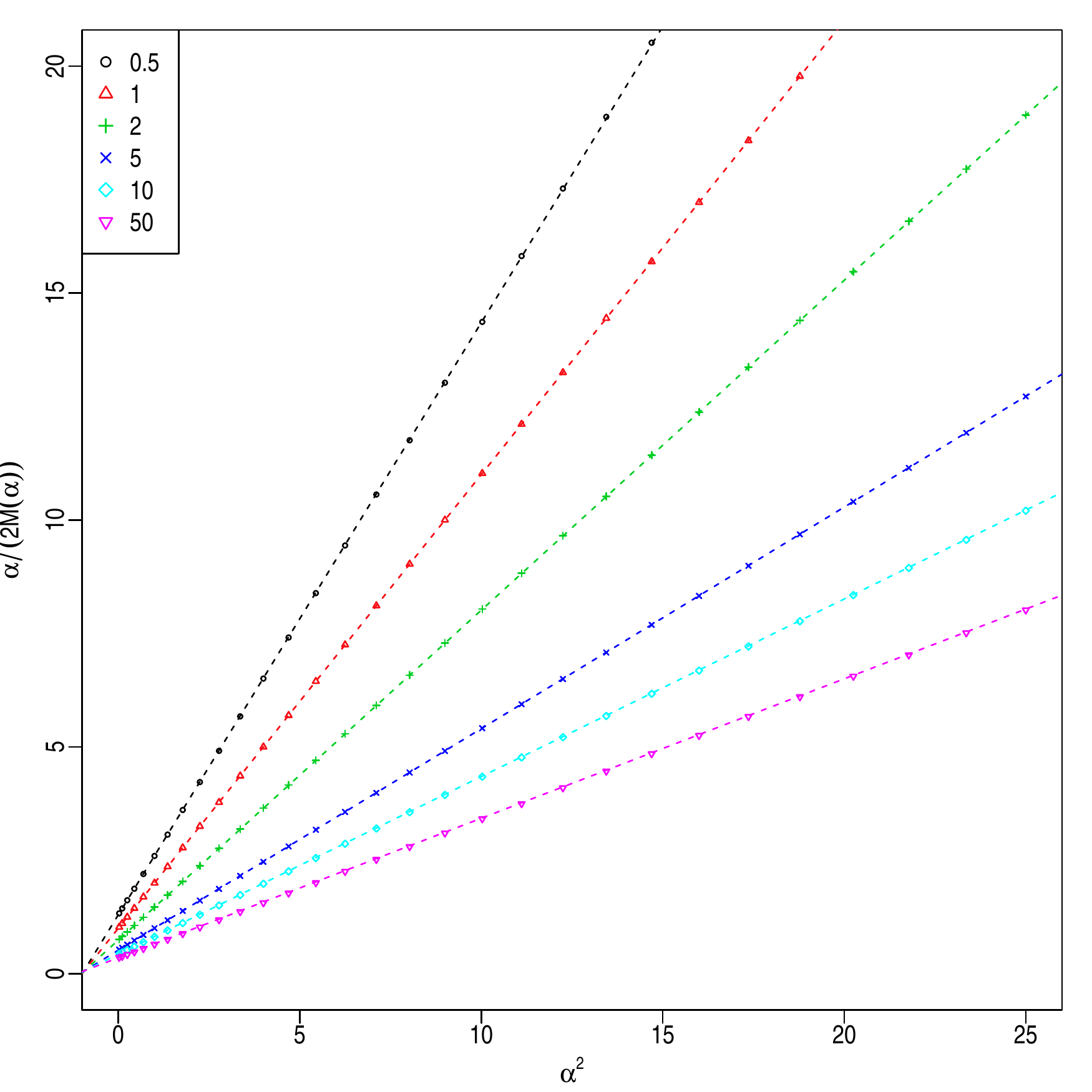}
   \hfill 
   \includegraphics[width=0.48\hsize]{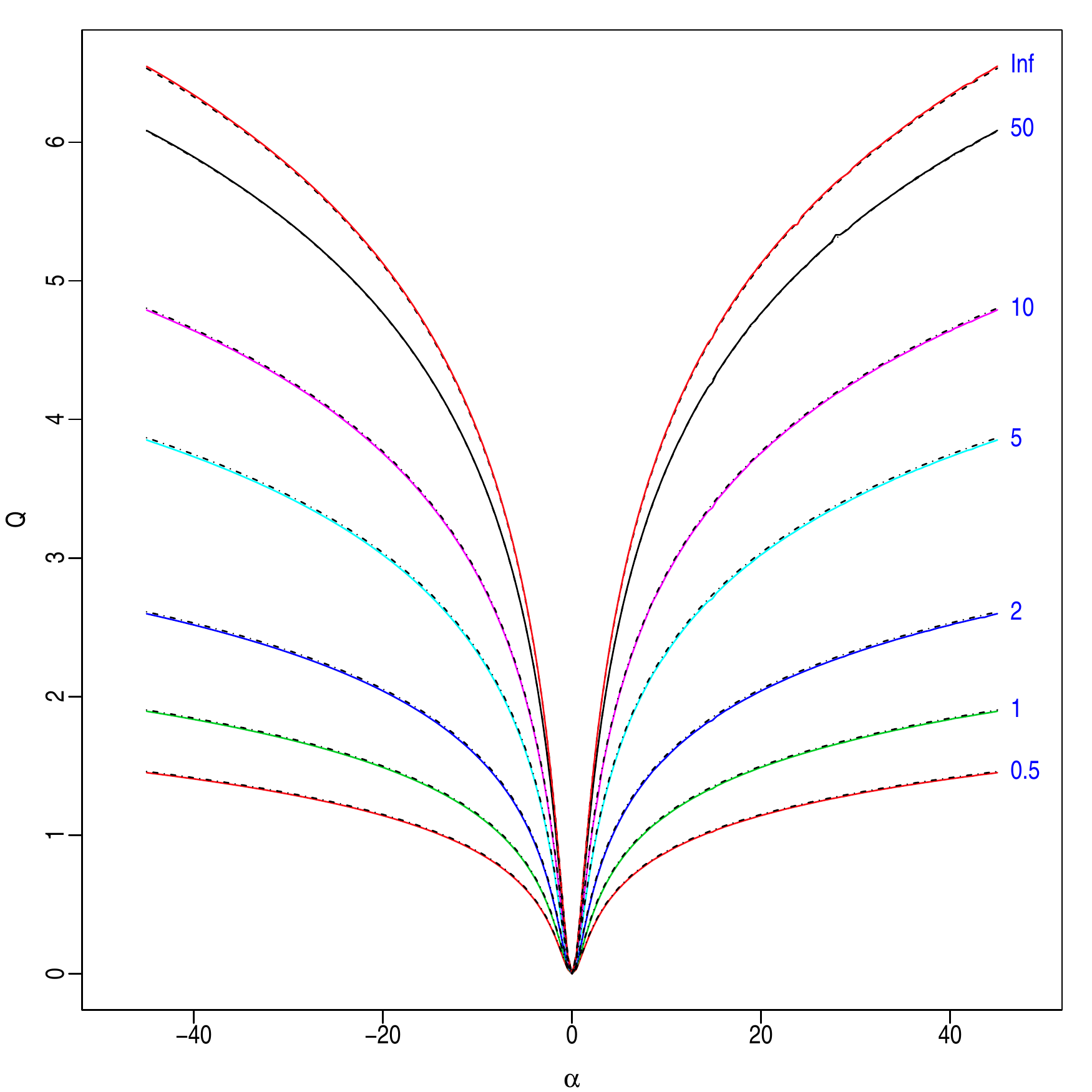}
   }
   \caption{\sl Left panel: values of $\alpha/[-2\,M(\alpha)]$ for ST
     distributions, for $\nu=\half, 1,2, 5, 10, 50$, numerically evaluated at
     a grid of points, plotted versus $\alpha^2$ and superimposed
     approximating line for each selected $\nu$.  
     Right panel: $Q$ function obtained by numerical integration of 
     $-M(\alpha)$ (continuous lines), and by $Q$ described in the text 
     (dot-dashed lines). }
\label{f:a2/a4&Q-ST}
\end{figure}

In the ST case, we proceed similarly to the SN case to obtain a close
approximation of the $M(\alpha)$ function. The plot on the left panel of
Figure~\ref{f:a2/a4&Q-ST} plays a similar role of the left plot in
Figure~\ref{f:a2/a4&Q-SN}, except that we now have a sequence of points for
each chosen values of $\nu$, specifically $\nu=\half, 1,2, 5, 10, 50$, with
different plotting symbols for each value of $\nu$. Also in this case there is
a striking alignment of the points referring to the same $\nu$, particularly
so if we consider the wide range of $\nu$ values considered.

For any fixed value of $\nu$, we approximate $\alpha/[-2\,M(\alpha)]$  
by a function of the form
\[  \frac{\alpha}{-2\,M(\alpha)} \approx e_{1\nu}+ e_{2\nu}\,\alpha^2 \]
whose coefficients $e_{1\nu}$ and $e_{2\nu}$ are obtained by matching
the behaviour of the two sides at $\alpha^2=0$ and $\alpha^2\to\infty$.
After some algebraic work summarized in an appendix, we arrive at
the expressions
\begin{equation}  \label{e:e1e2-ST}
  \begin{array}{rcl}
  e_{1\nu} &=& \dfrac{g_\nu}{3}\,,\\
  e_{2\nu} &=& g_\nu^2\:
       \dfrac{\E{X_1^2 \zeta_1(X_1;\nu+1)}}%
         {\E{X_3^4\,\zeta_1\left(X_3\sqrt{(\nu+1)/(\nu+3)}; \nu+1\right)}}
\end{array}    
\end{equation}  
where 
\begin{equation}  \label{e:zeta1}
   g_\nu = \dfrac{(\nu+2)\,(\nu+3)}{(\nu+1)^2} \,, \qquad
   \zeta_1(x;\nu) = \frac{t(x;\nu)}{T(x;\nu)}  \,.
\end{equation}
and $X_k\sim t_{\nu+k}$. The two expected values involved by $e_{2\nu}$ must
be evaluated numerically. The dashed lines in the left panel of
Figure~\ref{f:a2/a4&Q-ST} have intercepts and slopes given by \ref{e:e1e2-ST};
it is apparent that the lines interpolate the points described earlier almost
exactly.

Again, the integral of $-M(\alpha)$ is then closely approximated by
\ref{e:Q-SN1} with coefficients $c_1=1/(4\,e_{2\nu})$ and
$c_2=e_{2\nu}/e_{1\nu}$. The right panel of Figure~\ref{f:a2/a4&Q-ST} displays
the $Q$ function obtained by numerical integration of $-M(\alpha)$ as
continuous lines, and their approximation of the form \ref{e:Q-SN1} as
dot-dashed lines. The two set of lines are in fact virtually
indistinguishable. Therefore this choice of $Q$ is essentially equivalent to
the one of \citet{sartori:2006} and \cite{lago:jime:2012} in the case
$\ST(0,1,\alpha,\nu)$ when $\nu$ is regarded as known and only $\alpha$ is
estimated, but it has the computational advantage of avoiding the numerical
evaluation of a very large number of integrals, which are required when the
numerical algorithm for solving $M(\alpha)=0$ searches over a range of
$\alpha$ values.

At variance from the above-quoted authors, we adopt the penalty function just
described also in the four parameter case $\ST(\xi,\omega^2,\alpha, \nu)$,
analogously to what we did for the three-parameter SN case.  A point of
practical concern in this process is that, when $\nu$ is not fixed, the
algorithm for numerical optimization of $\ell_p(\theta)$ visits many candidate
values of $\nu$, and each of them involves numerical evaluation of the two 
integrals involved by $e_{2\nu}$ in \ref{e:e1e2-ST}.  
To avoid these extensive integrations, we have explored a simple empirical 
approximation of $e_{2\nu}$. 

\begin{figure}
\centerline{
   \includegraphics[width=0.48\hsize]{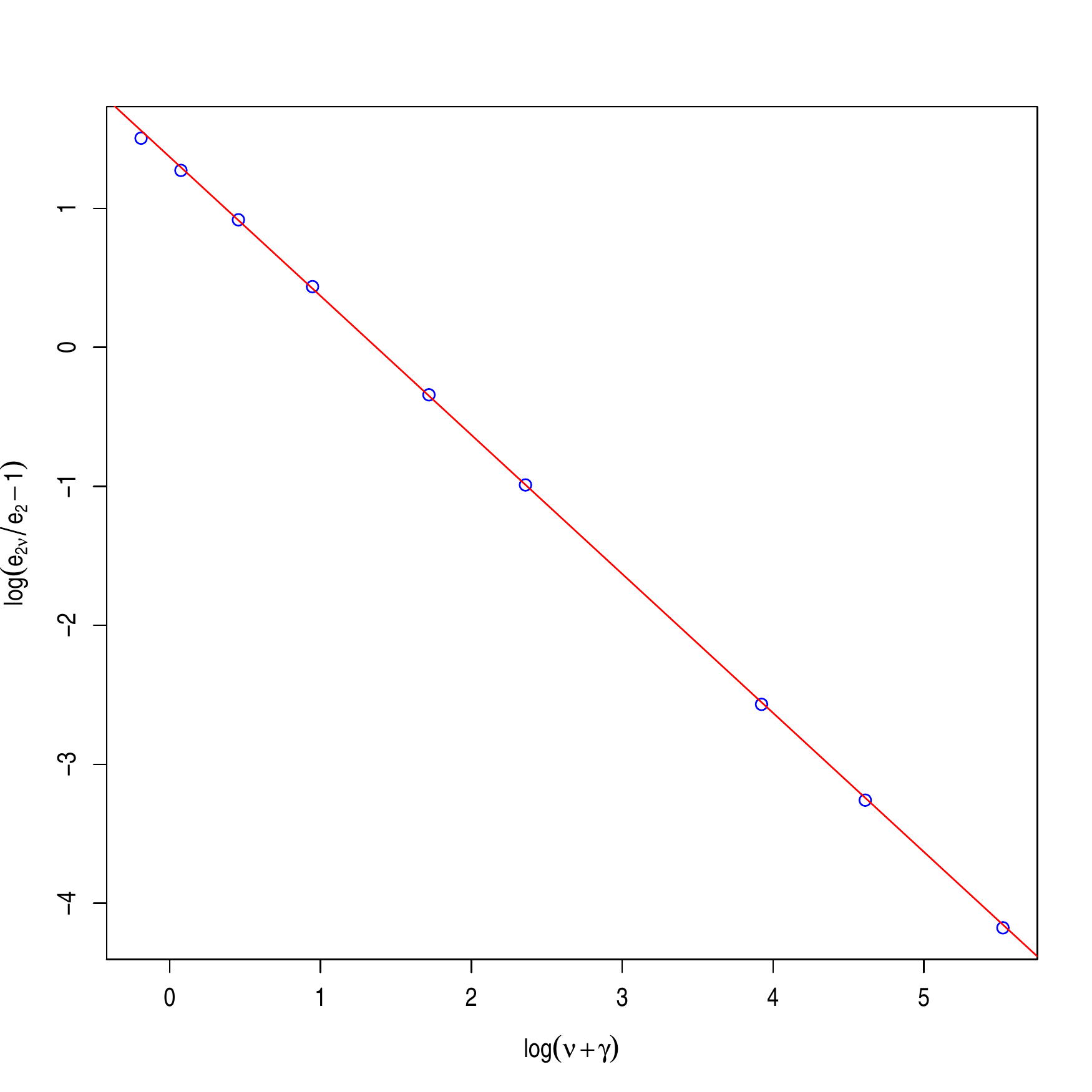}
   \hfill
   \includegraphics[width=0.48\hsize]{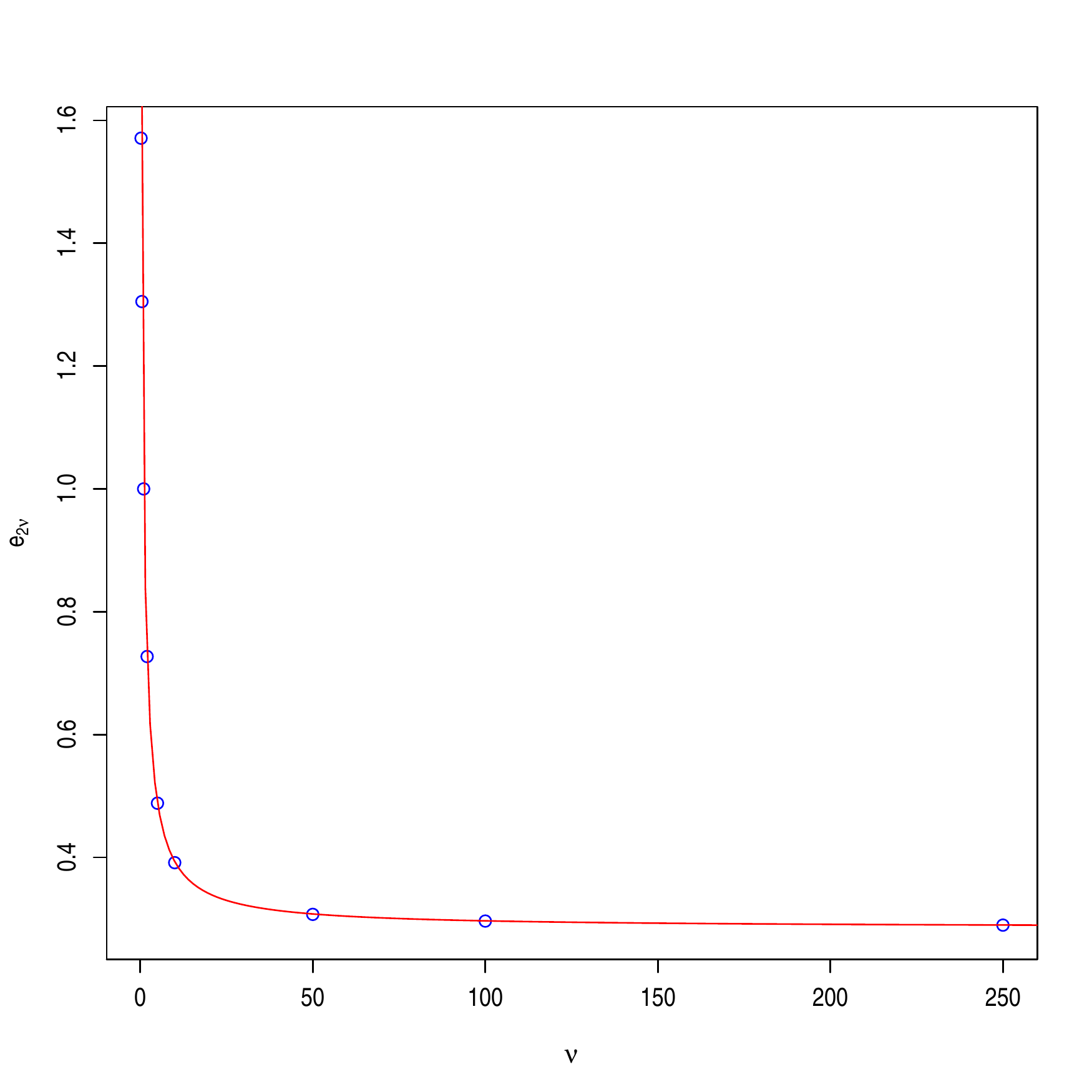}
   }
   \caption{\sl Exact values of $e_{2\nu}$  and  their 
   approximating function on the transformed scale (left side plot) 
   and on the  original scale (right side plot).}
\label{f:interp-e2-ST}
\end{figure}

Some numerical exploration has show that $\log(e_{2\nu}/e_2 -1)$ is 
very nearly a linear function of $\log(\nu+\gamma)$ where $
\gamma=0.57721\dots$ is the Euler's constant, 
and $e_2$ is the limiting value \ref{e:e1e2-SN}.  
This near linearity is visible in the left plot of 
Figure~\ref{f:interp-e2-ST} 
which displays a set of numerically evaluated values $e_{2\nu}$, 
for a range of degrees of freedom from $\nu=0.25$ to $\nu=250$, 
transformed to $\log(e_{2\nu}/e_2 -1)$ and plotted versus 
$\log(\nu+\gamma)$. 
The interpolating line fitted by least squares has intercept $1.37$ and slope
$-1.00$ when rounded to two decimal places; these are the coefficients of the
plotted line. In the right-side plot of the figure, the interpolation has
been transformed back on the original scale, so that the continuous line
superimposed to the points $(\nu, e_{2\nu})$ is the approximation
\[ 
  e_{2\nu} \approx e_2\left(1+ \frac{4}{\nu+\gamma}\right) 
\] 
which appears to work well.   Clearly the other coefficient,
$e_{1\nu}=g_\nu/3$, poses no problem.

In the multivariate case we proceed similarly to the skew-normal case, and
consider a penalized log-likelihood function similar to \ref{e:logL.p-SN},
but the coefficients $c_1$ and $c_2$ now depend on $\nu$, and the
summands of the first term are replaced by the logarithm of the multivariate
skew-$t$ density function. This density is given by
\begin{equation} \label{e:dmst}
    2\, t_d(x-\xi; \Omega, \nu)\:
    T\left(\sqrt{\frac{d+\nu}{Q_x+\nu}}\,
         \alpha\T\omega\inv(x-\xi); \, \nu+d\right),
   \qquad x\in\Real^d,
\end{equation}  
where $t_d(x;\Omega)$ denotes the $d$-dimensional Student's $t$ density
function with location $0$, scale matrix $\Omega$ and $\nu$ degrees of
freedom; the earlier definition of $Q_x$ is now replaced by 
$Q_x=(x-\xi)\T\Omega\inv (x-\xi)$.

% ======================================================
\section{Discussion}
 
For the SN and ST distributions in the univariate and multivariate cases, we
have examined a methodology which avoids the problem of estimates on the
frontier of the parameter space which can occur with maximum likelihood
estimation. The problem of estimates on the frontier is vanishing when the
sample size diverges, but in practice, for small to moderate sample sizes, it
arises with non-negligible probability and it has disturbing effects on the
inferential process.

The present proposal is, in a way, closely connected with existing results,
but there are differences. One is that the focus here is shifted from
unbiasedness to the use of a penalty function $Q$ which can be chosen quite
freely once a few requirements are satisfied. This is the basis of another
difference from existing work: we have adopted the $Q$ function arising from
the one-parameter cases, $\SN(0,1,\alpha)$ and $\ST(0,1,\alpha,\nu)$ with
fixed $\nu$, as the starting point for the choice of $Q$ in more complex
situations, that is multiparameter and multivariate settings.

There are various directions in which the present work can be extended,
of which we mention a few.
\begin{itemize}
\item As it stands, the methodology presented here is applicable to the
  skew-normal and the skew-$t$ distributions, which are the two most commonly
  employed families from the broader set of skew-elliptical distributions.
  However the formulation could be adapted to other skew-elliptical families.
\item The asymptotic results of Section~\ref{s:logL.p} are all of first-order
  type. There is wide room for higher asymptotic theory; accuracy of
  approximation \ref{e:var(MPLE)} is of special interest.
\item Alternative choices of the penalty $Q$ could be considered provided the
  new function satisfies conditions \ref{e:Q-assume} and those indicated
  shortly thereafter.
\end{itemize}
 
While these  developments are interesting, the proposal at its present stage 
provides an already workable and quite general way to overcome what appears 
to us as the last obstacle to the systematic use of skew-normal and skew-$t$ 
distributions in routine statistical work. 

% A number of papers have shown the practical usefulness of these 
% distributions in practical problems; see for instance 
% \citet{azza:gent:2008} for a fairly extensive illustration of this sort.

Before closing, it is perhaps useful to point up that the adoption
of the penalized likelihood formulation is compatible with the adoption of the
centred parameterization mentioned in the introductory section as a tool to
overcome the singularity of the information in the skew-normal case. The two
mechanisms are conceptually  distinct and they can coexist. Once the
MPLE of the direct parameters have been obtained, they can be transformed into
the centred parameter space, and the variance matrix of the MPLE estimates can
be converted via the known Jacobian matrix of the transformation
\citep{arel:azza:2008}. For the skew-$t$ distribution the problem of singular
information matrix at $\alpha=0$ does not arise for any $0<\nu<\infty$,
as proved in Proposition~1 of \citet{arellano-valle:2010}.
%\NB{cite Ley \& Hallin?}
% --------------------------------------------- 
\subsection*{Acknowledgement}
This work was initiated while the first author was visiting the 
Departamento de Estadística,  Pontificia Universidad Católica de Chile,
whose generous hospitality is gratefully acknowledged. 
The research work of the second author was partially supported 
by grant FONDECYT 1085241, Chile
% --------------------------------------------- 
%\small
%\bibliography{strings,sn+related,misc-distn,books,miscellanea,thispaper}

% --------------------------------------------- 
\appendix
\section*{Appendix: Limiting behaviour of $M(\alpha)$ in the ST case}
Consider a random variable  $Z\sim \ST(0,1,\alpha,\nu)$ and its
transformation  $W=v(Z)\,Z$ where
\[ v(z)=\sqrt{\frac{\nu+1}{\nu+z^2}}  \,. \]
From \cite{sartori:2006}, we write
\begin{eqnarray*}
 l'(\alpha)&=&\zeta_1(\alpha W;\nu+1)\,W,\\
 l''(\alpha)&=&-\frac{\nu+1}{\nu+2}
   \left(1+\frac{\alpha^2 W^2}{\nu+1}\right)\inv
    \zeta_1(\alpha W;\nu+1)\,W^3-\zeta_1(\alpha W;\nu+1)^2\,W^2,\\
 l'(\alpha)^3+l'(\alpha)\,l''(\alpha) &=&
   -\frac{\nu+1}{\nu+2}\left(1+\frac{\alpha^2 W^2}{\nu+1}\right)\inv
    \alpha\, \zeta_1(\alpha W;\nu+1)^2\,W^4.
 \end{eqnarray*}
where $\zeta_1(x;\nu)$ is the function defined by \ref{e:zeta1}.
Let now $X_k\sim t(0,1;\nu+k)$, and define the random variables
%$W_0=v(Z_0)Z_0$ and
\[V_{k\delta}=\sqrt{\frac{\nu+1}{\nu+k+(1-\delta^2)\,X_k^2}}.\]
After some simple algebraic manipulations, where we use the
relation
\begin{eqnarray*}
t(z;\nu)t\left(\alpha z\sqrt{\frac{\nu+1}{\nu+z^2}};\nu+1\right)
 &=&
   t(0;\nu)\sqrt{\frac{\nu+1}{\nu}}\,
  \left(\frac{\nu+1}{\nu+z^2}\right)^{-1/2}t\left(\sqrt
   {\frac{\nu+1}{\nu}}\,\sqrt{1+\alpha^2}\, z;\nu+1\right),
\end{eqnarray*}
we obtain
\begin{eqnarray*}
\E{l''(\alpha)}&=&-\E{\zeta_1(\alpha W;\nu+1)^2\,W^2}\\
&=&-(1+\alpha^2)^{-3/2}b_\nu\sqrt{\frac{\nu+1}{\nu}}\,
\E{X_1^2\,V_{1\delta}\zeta_1(\delta X_1\,V_{1\delta};\nu+1)},\\
\E{l'(\alpha)^3}+\E{l'(\alpha)\,l''(\alpha)}
&=&-\alpha\left(\frac{\nu+1}{\nu+2}\right)
  \E{\zeta_1(\alpha W;\nu+1)^2
     \left(1+\frac{\alpha^2 W^2}{\nu+1}\right)\inv \,W^4}\\
&=&-\alpha(1+\alpha^2)^{-5/2}b_\nu\sqrt{\frac{\nu}{\nu+3}}
\left(\frac{\nu+1}{\nu+2}\right)^2\left(\frac{\nu+1}{\nu+3}\right)\:\times \\
&&\qquad 
  \E{X_3^4 \,V_{3\delta}\, \zeta_1(\delta X_3\, V_{3\delta};\nu+1)}
\end{eqnarray*}
where $b_\nu=2t(0;\nu)$. Thus, using the Sartori--Firth formulae for
$M(\alpha)$, we write
\begin{eqnarray*}
M(\alpha)
&=&\frac{\E{l'(\alpha)^3}+\E{l'(\alpha)\,l''(\alpha)}}{-2\E{l''(\alpha)}}\\
&=&-\frac{\alpha}{2}\,\frac{\left(\frac{\nu+1}{\nu+2}\right)
{\small\E{\zeta_1(\alpha W;\nu+1)^2
 \left(1+\frac{\alpha^2\, W^2}{\nu+1}\right)\inv W^4}}}
  {\E{\zeta_1(\alpha W;\nu+1)^2\,W^2}}\\
&=&-\frac{\alpha(1+\alpha^2)^{-5/2}b_\nu\sqrt{\frac{\nu}{\nu+3}}
 \left(\frac{\nu+1}{\nu+2}\right)^2\left(\frac{\nu+1}{\nu+3}\right)\,
 \E{X_3^4\,V_{3\delta}\,\zeta_1(\delta X_3\, V_{3\delta};\nu+1)}}
 {2(1+\alpha^2)^{-3/2}b_\nu\sqrt{\frac{\nu}{\nu+1}}\,
  \E{X_1^2\,V_{1\delta}\,\zeta_1(\delta X_1\,V_{1\delta};\nu+1)}}.
\end{eqnarray*}
Note that the second expression agrees with a matching one of
\cite{lago:jime:2012}. From the last expression, we obtain
\begin{eqnarray*}
-\frac{\alpha}{2\,M(\alpha)}
&=& (1+\alpha^2) \left(\frac{\nu+2}{\nu+1}\right)^2
\left(\frac{\nu+3}{\nu+1}\right)^{3/2}
\frac{\E{X_1^2\,V_{1\delta}\,\zeta_1\left(\delta X_1\, V_{3\delta};\nu+1\right)}}
{\E{X_3^4\,V_{3\delta}\, \zeta_1\left(\delta X_3\,V_{3\delta};\nu+1\right)}}\\
&\approx& e_{1\nu}+e_{2\nu}\alpha^2.
\end{eqnarray*}
Thus, by noting  that for $\alpha^2=0$
\begin{eqnarray*}
\E{X_k^{2r}\,V_{k0}}&=&\frac{b_{\nu+k}}{b_{\nu+k+1}}
   \sqrt{\frac{\nu+1}{\nu+k}}
  \left(\frac{\nu+k}{\nu+k+1}\right)^{(2r+1)/2}\E{X_{k+1}^{2r}}\\
  &=& \left(\frac{b_{\nu+k}}{b}\right)^2\sqrt{\frac{\nu+1}{\nu+k}}
  \left(\frac{\nu+k}{2}\right)^r
  \frac{\Gamma[(\nu+k+1-2r)/2]}{\Gamma[(\nu+k+1)/2]}\frac{(2r)!}{2^r r!},
\end{eqnarray*}
we arrive at
\begin{eqnarray*}
e_{1\nu}&=&\left(\frac{\nu+2}{\nu+1}\right)^2
  \left(\frac{\nu+3}{\nu+1}\right)^{3/2}\frac{\E{X_1^2V_{10}}}
  {\E{X_3^4V_{30}}}\\
&=&\frac{1}{3} \left(\frac{b_{\nu+1}}{b_{\nu+3}}\right)^2
  \left(\frac{\nu+2}{\nu+1}\right)^3,
\end{eqnarray*}
while
\begin{eqnarray*}
  e_{2\nu}&=&\lim_{\alpha^2\to\infty}\left\{\frac{1+\alpha^2}{\alpha^2}
   \left(\frac{\nu+2}{\nu+1}\right)^2
      \left(\frac{\nu+3}{\nu+1}\right)^{3/2}
    \,\frac{\E{X_1^2V_{1\delta} \zeta_1\left(\delta
          X_1V_{1\delta};\nu+1\right)}}{\E{X_3^4V_{3\delta} \zeta_1\left(\delta
          X_3V_{3\delta};\nu+1\right)}}-\frac{e_{1\nu}}{\alpha^2}\right\}\\
  &=&\left(\frac{\nu+2}{\nu+1}\right)^2
    \left(\frac{\nu+3}{\nu+1}\right)^{2}\,
    \frac{\E{X_1^2 \,\zeta_1( X_1;\nu+1)}}%
     {\E{X_3^4 \,\zeta_1\left(\sqrt{\nu+1}\,X_3/\sqrt{\nu+3};\nu+1\right)}}.
\end{eqnarray*}
\end{document}